\documentclass[fleqn,usenatbib]{mnras}

\usepackage{newtxtext,newtxmath}

\usepackage[T1]{fontenc}
\usepackage{ae,aecompl,times}

\usepackage{graphicx}	
\usepackage{amsmath}	
\usepackage{amssymb}	
\usepackage{multirow}
\usepackage[normalem]{ulem}

\title[Marked correlations in $f(R)$ gravity]{Marked clustering statistics in $f(R)$ gravity cosmologies}

\author[Hern\'andez-Aguayo, Baugh, Li]{
C\'esar Hern\'andez-Aguayo,$^{1}$\thanks{E-mail: cesar.hernandez-aguayo@durham.ac.uk (ICC)}
Carlton M. Baugh,$^{1}$
Baojiu Li$^{1}$
\\
$^{1}$Institute for Computational Cosmology, Department of Physics, Durham University, South Road, Durham DH1 3LE, UK \\
}

\date{Accepted XXX. Received YYY; in original form ZZZ}

\pubyear{2018}

\begin{document}
\label{firstpage}
\pagerange{\pageref{firstpage}--\pageref{lastpage}}
\maketitle

\begin{abstract}
We analyse the two-point and marked correlation functions of haloes and galaxies in three variants of the chameleon $f(R)$ gravity model using N-body simulations, and compare to a fiducial $\Lambda$CDM model based on general relativity (GR). Using a halo occupation distribution prescription (HOD) we populate dark matter haloes with galaxies, where the HOD parameters have been tuned such that the galaxy number densities and the real-space galaxy two-point correlation functions in the modified gravity models match those in GR to within $1\sim3\%$. We test the idea that since the behaviour of gravity is dependent on environment, marked correlation functions may display a measurable difference between the models. For this we test marks based on the density field and the Newtonian gravitational potential. We find that the galaxy marked correlation function shows significant differences measured in different models on scales smaller than $r\lesssim 20~h^{-1}$ Mpc. Guided by simulations to identify a suitable mark, this approach could be used as a new probe of the accelerated expansion of the Universe. 
\end{abstract}





\section{Introduction}
The $\Lambda$ cold dark matter ($\Lambda$CDM) model is currently the most widely accepted description of the Universe (e.g. see \citealt{Ade:2013zuv}). In this model small ripples in the density of the Universe at early times, seeded during a period of rapid expansion called inflation, were boosted by gravity to form the cosmic web of voids, galaxies and clusters of galaxies that we see today. 
The $\Lambda$CDM model works remarkably well on large scales, with the highlight being the prediction of the temperature fluctuations in the cosmic microwave background (CMB) radiation (e.g. \citealt{Planck:cosmoparams}). However, the model arguably runs into difficulties on small scales, which could be related to the nature of the dark matter particle or could be solved by appealing to the physics of galaxy formation (for a review see \citealt{Weinberg:2013aya}). 

Several large international survey projects are underway which aim to determine what is behind the accelerating expansion of the Universe, such as the Dark Energy Survey (DES)~\citep{DES:2016}, the Dark Energy Spectrographic Instrument (DESI) survey~\citep{DESI:2016} and the European Space Agency's Euclid mission~\citep{Laureijs:2011gra}. These surveys aim to make ambitious measurements which, for the first time, will be dominated by systematics rather than by the volume of the Universe surveyed. Accurate theoretical predictions are urgently required to compare with these expected measurements and to rule out competing scenarios for the accelerating cosmic expansion.

Theoretically, the $\Lambda$CDM model is somewhat unappealing due to the presence of the cosmological constant, the agent behind the accelerating cosmic expansion. The magnitude of the cosmological constant is hard to motivate from a particle physics perspective. There is also a ``Why now?'' problem: a strong coincidence seems to be required for us to be at the right point in cosmic history to experience comparable energy densities in matter and the cosmological constant, with the latter dominating the current expansion. As a result, alternatives to the cosmological constant have been studied extensively in recent years: one possibility is adding more matter species to the energy-momentum tensor (the so-called dark energy models; see e.g., \citealt{Copeland:2006wr}); on the other hand there are models that change the left-hand side of Einstein's equations (these models are called modified gravity models, for reviews see \citealt{Joyceetal2015,Koyama2016}). 

Here, we focus our attention on a particular class of modified gravity models -- \citet{Hu:2007nk} chameleon $f(R)$ gravity. This model is obtained by adding a general function of the Ricci scalar, $f(R)$, to the Einstein-Hilbert action. This modification gives rise to a new scalar degree of freedom in gravity \citep{Carroll:2003wy}. In order to recover GR, the new scalar field becomes massive in high-density regions (i.e., the Solar System) and its interactions are suppressed by the so-called chameleon screening mechanism \citep{Khoury:2003rn}. 

The standard tool to model the growth of large scale structure into the non-linear regime is N-body simulation. In order to robustly test gravity on cosmological scales, reliable N-body simulations of modified gravity models are essential. The non-linear nature of the scalar field equation requires the implementation of novel numerical techniques, which is what makes N-body simulations of modified gravity challenging \citep{Winther:2015wla,Barreira:2015xvp,Bose:2016wms}. Once such simulations are ready, we can use measurements of clustering statistics from surveys to test and constrain cosmological models (see e.g., \citealt{Reid:2009xm}).

Several recent works have studied clustering in $f(R)$ gravity. For example, \citet{Li:2012by} predicted the matter and velocity divergence power spectra and their time evolution measured from several large-volume N-body simulations with varying box sizes and resolution. \citet{Jennings:2012pt} predicted the clustering of dark matter in redshift space, finding significant deviations from the clustering signal in standard gravity, with an enhanced boost in  power on large scales and stronger damping on small scales in the $f(R)$ models compared to GR at redshifts $z<1$. More recently, \citet{Arnalte-Mur:2016alq} compared the time evolution of the two-point correlation function of dark matter haloes in real and redshift space in modified gravity and GR. 

An approach related to the two-point correlation function has been proposed to test modified gravity models, called the marked  correlation function \citep*{Sheth:2005aj}.  Marked statistics offer the possibility of testing how galaxy properties correlate with environment. Previous applications of the marked correlation function range from the analysis of the environmental dependence of bars and bulges in disc galaxies (see e.g., \citealt{Skibba:2011ax}) to breaking degeneracies in halo occupation distribution modelling \citep{White:2008ii}.

\citet{White:2016yhs} proposed that marked statistics might provide a means to distinguish between modified gravity models and GR, due to the environmental dependence of the strength of gravity (for earlier work  on environmental dependence in modified gravity, see, e.g., \citealp{Zhao:2011cu,Winther:2011qb,Lombriser:2015axa,Shi:2017pyd}). Marks can be designed which down-weight high-density regions, for which modifications to gravity are screened, and up-weight low-density, unscreened regions to maximise the differences in the clustering signal. \citet{Valogiannis:2017yxm} tested this idea by using the dark matter particle distribution from N-body simulations of symmetron and $f(R)$ modified gravity models. In the case of $f(R)$ gravity with $|f_{R0}| = 10^{-4}$, they found a maximum difference of $37\%$ with respect to GR. \cite{Joaquin:2018} studied the galaxy marked correlation function by up-weighting low and high-density regions using marks in function of the galaxy density field and the host halo mass of galaxies, they found significant differences between the $f(R)$ and GR models.  Here, we focus our attention on the clustering of dark matter haloes and HOD galaxies for $f(R)$ gravity models to make a more direct connection with observations.  

This paper is organized as follows. In Section~\ref{sec:f(R)} we give a brief review of the theoretical description of $f(R)$ gravity and the physical motivation behind this model. Section~\ref{sec:simulations} explains the numerical set-up of the simulations and the generation of halo and galaxy catalogues. The main results are presented in Section~\ref{sec:halo}. Finally, in Section~\ref{sec:conclusions} we present a brief discussion and our general conclusions.

\section{\lowercase{$f$}$(R)$ gravity theory}\label{sec:f(R)}
A popular family of modified gravity models is obtained by replacing the Ricci scalar $R$ in the usual Einstein-Hilbert Lagrangian density by some function $f(R$) (see~\citealt{Sotiriou:2008rp,DeFelice:2010aj}
for recent reviews). These models are often considered as an alternative solution for the accelerating cosmic expansion. However, it should be noted that they are able to accelerate the expansion only because a cosmological constant is added ``through the back door'' \citep{Brax:2008hh,Wang:2012kj,Ceron-Hurtado:2016jrp}, and so they do not offer a real solution to the cosmological constant problem. Nevertheless, they constitute a  representative model with which to study cosmological constraints on possible deviations from GR.

\subsection{Theoretical framework}\label{sub:T-f(R)}
The action of $f(R)$ theories is given by~\citep{Carroll:2003wy}:
\begin{equation}\label{eq:S-f(R)}
S = \int \mathrm{d}^4x\sqrt{-g} \frac{1}{16\pi G} (R + f(R)) + S_{\rm m}(g_{\mu\nu},\psi_i)\, ,
\end{equation}
where $G$ is Newton's constant, $g$ is the determinant of the metric $g_{\mu\nu}$ and $S_{\rm m}$ is the action of the matter fields $\psi_i$ (including the contributions from cold dark matter, baryons, radiation and neutrinos). 

Varying the action given in Eqn.~\eqref{eq:S-f(R)}, with respect to the metric, $g_{\mu\nu}$, one obtains the modified Einstein equations
\begin{equation}\label{eq:M-EQ}
G_{\mu\nu} + f_R R_{\mu\nu} - g_{\mu\nu} \left(\frac{1}{2}f(R) - \Box f_R \right) - \nabla_\mu \nabla_\nu f_R = 8\pi G T^{\rm m}_{\mu\nu}\,,
\end{equation}
where $G_{\mu\nu} = R_{\mu\nu} - \frac{1}{2}g_{\mu\nu} R$ is the Einstein tensor, $\nabla_\mu$ is the covariant
derivative, $\Box = \nabla^\mu \nabla_\mu$ the d'Alambertian, and $T^{\rm m}_{\mu \nu}$ is the energy-momentum tensor for matter.
Here
\begin{equation}\label{eq:sc}
f_R \equiv \frac{\mathrm{d}f(R)}{\mathrm{d}R} , 
\end{equation}
is the extra degree of freedom of this model, known as the \emph{scalaron} field.

Taking the trace of Eqn.~\eqref{eq:M-EQ}, we obtain the equation of motion for the scalaron field,
\begin{equation}\label{eq:f_R}
\Box f_R = \frac{1}{3} (R - f_R R + 2f(R) + 8\pi G\rho_{\rm m})\,,
\end{equation}
where $\rho_{\rm m}$ is the matter density of the Universe.

Since we are interested in the cosmological evolution of the model, we derive the perturbation equations using the flat Friedmann-Robertson-Walker (FRW) metric in the Newtonian gauge
\begin{equation}\label{eq:FRW}
\mathrm{d}s^2 = (1+2\Psi) \mathrm{d}t^2 - a^2(t)(1-2\Phi)\gamma_{ij}\mathrm{d}x^i \mathrm{d}x^j\,,
\end{equation}
where $\Psi$ and $\Phi$ are the gravitational potentials, $t$ is the cosmic time, $x^i$ represent the comoving coordinates, $\gamma_{ij}$ is the 3D metric, and $a$ is the scale factor, with $a(t_0) = a_0 = 1$ at the present time.

In the quasi-static and weak field limits \citep*{Bose:2014zba}, structure formation in this model is determined by the following equations:
\begin{equation}\label{eq:Phi}
\nabla^2 \Phi = \frac{16\pi G}{3} a^2 \delta \rho_{\rm m} + \frac{1}{6} a^2 \delta R\,,
\end{equation}
\begin{equation}\label{eq:fR}
\nabla^2 f_R = -\frac{a^2}{3} [\delta R + 8\pi G\delta \rho_{\rm m}]\,,
\end{equation}
in which $\nabla^2$ is the three-dimensional Laplacian operator, 
$\delta \rho_{\rm m} = \rho_{\rm m} - \bar{\rho}_{\rm m}$ and $\delta R = R(f_R) - \bar{R}$ are, 
respectively, the density and Ricci scalar perturbations (overbars denote 
background cosmological quantities). 
\subsection{The chameleon mechanism}\label{sub:cham}
In $f(R)$ gravity the modifications to Newtonian gravity can be considered as a fifth force mediated by the scalaron field, $f_R$. The chameleon mechanism (see e.g., \citealt{Khoury:2003rn,Mota:2006fz}) was introduced to give scalar fields an environment-dependent effective mass allowing the scalar mediated force to be suppressed under certain environmental conditions. 

Because the scalaron field itself is massive, this fifth force is of Yukawa type, i.e. decaying exponentially as $\exp (-m_\mathrm{eff} r)$, where $m_\mathrm{eff}$ is the scalaron mass
\begin{equation}\label{eq:m_eff}
m^2_\mathrm{eff} \equiv \frac{\mathrm{d}^2V_\mathrm{eff}}{\mathrm{d} f^2_R}\,,
\end{equation}
and the effective potential is related to the trace of the modified Einstein Eqn.~\eqref{eq:f_R}
\begin{equation}\label{eq:V_eff}
\frac{\mathrm{d}V_\mathrm{eff}}{\mathrm{d} f_R} = \frac{1}{3} (R - f_R R + 2f(R) + 8\pi G\rho_{\rm m})\,.
\end{equation}
In $f(R)$ gravity this fifth force enhances gravity in regions of weak gravitational potential and on scales below the Compton wavelength of the scalar field, $\lambda = m_\mathrm{eff}^{-1}$.
While in high density regions, $m_\mathrm{eff}$ is heavy and the fifth force is more strongly suppressed and gravity reverts to GR.
\subsection{The Hu-Sawicki model}\label{sub:HS}
We consider the \citeauthor{Hu:2007nk} model:
\begin{equation}\label{eq:f(R)}
f(R) = -m^2 \frac{c_1}{c_2} \frac{(-R/m^2)^n}{(-R/m^2)^n + 1}\,,
\end{equation}
where $m$ is a characteristic mass scale, defined by $m^2 = 8\pi G \bar{\rho}_{\rm m0}/3 = H^2_0 \Omega_{\rm m}$, $\Omega_{\rm m}$ is the cosmological matter density parameter, $H_0$ is the present-day value of the Hubble parameter, and $n$, $c_1$ and $c_2$ are dimensionless parameters of the model. The scalaron field, Eqn.~\eqref{eq:sc}, takes the form:
\begin{equation}\label{eq:sc1}
f_R = -\frac{c_1}{c^2_2} \frac{n(-R/m^2)^{n-1}}{[(-R/m^2)^n + 1]^2}\,.
\end{equation}

To match the background expansion to that in the $\Lambda$CDM model, we set
\begin{equation}\label{eq:c1c2}
\frac{c_1}{c_2} = 6\frac{\Omega_\Lambda}{\Omega_{\rm m}}\,,
\end{equation}
where $\Omega_\Lambda \equiv 1 - \Omega_{\rm m}$. 

The expression of the scalaron field, Eqn.~\eqref{eq:sc1}, simplifies when the background value of the Ricci scalar satisfies $|\bar{R}| \gg m^2$. From
\begin{equation}\label{eq:R}
-\bar{R} \approx 8\pi G\bar{\rho}_{\rm m} - 2f(\bar{R}) \approx 3 m \left[ a^{-3}  + \frac{2}{3} \frac{c_1}{c_2}\right]\,,
\end{equation}
when $\Omega_\Lambda \approx 0.7$ and $\Omega_{\rm m} \approx 0.3$ we find $|\bar{R}| \approx 40 m^2 \gg m^2$.
Thus
\begin{equation}\label{eq:sc2}
f_R \approx -n\frac{c_1}{c^2_2} \left(\frac{m^2}{-R} \right)^{n+1}\,.
\end{equation}

The model then has two remaining free parameters, $n$ and $c_1/c^2_2$. The latter is related to the present-day value of the background scalaron, $f_{R0}$,
\begin{equation}\label{eq:c12}
\frac{c_1}{c^2_2} = -\frac{1}{n} \left[ 3\left(1 + 4\frac{\Omega_\Lambda}{\Omega_{\rm m}} \right) \right]^{n+1} f_{R0}\,.
\end{equation}
Hence, the choice of $f_{R0}$ and $n$ fully specifies the model. Here we focus on the case of $n=1$, which is the most well-studied case in the literature.
\section{Simulations and halo/galaxy catalogues}\label{sec:simulations}

\begin{table}
\centering
\caption{Numerical parameters of the simulations used.}
\label{tab:sim}
\begin{tabular}{ll}
\hline \hline
Labels & GR, F6, F5, F4 \\
Present value of the scalaron field & $|f_{R0}| = 0, 10^{-6}, 10^{-5}, 10^{-4}$\\
Box size & $L_\mathrm{box} = 1024~h^{-1}\mathrm{Mpc}$ \\
Number of DM particles & $N_{\rm p} = 1024^3$ \\
Mass of DM particle & $m_{\rm p} = 7.798 \times 10^{10} h^{-1}M_\odot$ \\
Initial redshift & $z_\mathrm{in} = 49$ \\
Final redshift & $z_\mathrm{fi} = 0$ \\
Realisations & $5$ \\
\hline 
\emph{Cosmological parameters:} & \\
Total matter density & $\Omega_\mathrm{m} = 0.281$\\
$1 - \Omega_\mathrm{m}$ & $\Omega_\Lambda = 0.719$ \\
Baryonic matter density & $\Omega_\mathrm{b} = 0.046$ \\
Cold dark matter density & $\Omega_\mathrm{cdm} = 0.235$ \\
Dimensionless Hubble parameter & $h = 0.697$ \\
Primordial power spectral index & $n_s = 0.971$ \\
rms linear density fluctuation & $\sigma_8 = 0.820$ \\ \hline
\end{tabular}
\end{table}

Here we present a description of the simulations used, the construction of halo catalogues, and the HOD prescription used to populate dark matter haloes with galaxies. 

\subsection{Numerical simulations}\label{sub:num}
As we are interested in the effects of $f(R)$ gravity on large scales, we choose three Hu-Sawicki models with $n=1$ and $|f_{R0}| = 10^{-6},10^{-5},10^{-4}$ (which we hereafter refer to as F6, F5 and F4, respectively) and the $\Lambda$CDM model which assumes GR. Despite the observational tensions faced by $f(R)$ models with $|f_{R0}| > 10^{-5}$ (see e.g., \citealt{Lombriser:2014dua,Cataneo:2014kaa,Liu:2016xes}) it is interesting to consider a wide range of $f(R)$ models to study their impact on the halo/galaxy clustering.

We use the \textsc{elephant} (Extended LEnsing PHysics using ANalaytic ray Tracing) simulations executed using the code \textsc{ecosmog} \citep{Li:2011vk}, which is based on the adaptive mesh refinement ({\sc amr}) N-body code \textsc{ramses} \citep{Teyssier:2001cp}. Table~\ref{tab:sim} lists the properties of the simulations used in our analysis. The cosmological parameters were adopted from the best-fitting values to the WMAP 9 year CMB measurements \citep{Hinshaw:2012aka}. All simulations use $N_{\rm p} = 1024^3$ particles with a mass of $m_{\rm p}=7.798\times 10^{10} h^{-1} M_\odot$ to follow the evolution of the dark matter distribution in a volume of $V_\mathrm{box} = (1024~h^{-1}\mathrm{Mpc})^3$. The initial conditions were generated at $z_\mathrm{ini}=49$ using the {\sc MPgrafic} code \citep{Prunet:2008fv}. All simulations were run using the same initial conditions up to the present time, $z_\mathrm{fi} = 0$, generating $37+1$ snapshots. Here, we analyse the outputs at $z=0.5$. 

\subsection{Halo catalogues and mass function}\label{sub:halo_c}

Dark matter haloes are the building blocks of large-scale structure and the hosts of galaxies. Therefore, the study of their statistical properties, such as their abundance and clustering, is of great importance in understanding the nature of gravity. 
The halo catalogues were produced using the \textsc{rockstar} halo finder code \citep{Behroozi:2011ju}. {\sc rockstar} calculates halo masses using the spherical overdensity (SO) approach \citep{Cole:1996nfw}, including all  particles and substructures in the halo. We keep only the `parent' halo, omitting other substructures from our analysis.

We define the mass of a halo as $M_{200c}$, the mass within a sphere of radius $r_{200c}$,  which is the radius within which the mean overdensity is 200 times the critical density of the universe $\rho_c$,
\begin{equation}\label{eq:M200c}
M_{200c} = \frac{4\pi}{3} 200\rho_c r^3_{200c}\,.
\end{equation} 
 
The dark matter halo mass function (HMF) quantifies the number density of dark matter haloes as a function of their mass. The HMF is sensitive to the cosmological parameters, $\Omega_{\rm m}$, $\Omega_{\Lambda}$, and $\sigma_{8}$, and to modifications to gravity.
The $\Lambda$CDM model predicts an HMF in which the number of haloes increases with decreasing halo mass. $f(R)$ models predict more haloes than the $\Lambda$CDM model at almost all masses due to the enhancement of gravity. 
Theoretically, the halo mass function is given by \citep{Press:1973iz}
\begin{equation}
\frac{{\rm d} n}{{\rm d} M_{200c}} = f(\sigma) \frac{\bar{\rho}_{\rm m}}{M^2_{200c}} \left| \frac{{\rm d}\ln \sigma^{-1}}{{\rm d}\ln M_{200c}} \right|\,,
\end{equation}
where $\sigma$ is the linear theory variance in the matter perturbation, $\bar{\rho}_{\rm m}$ is the mean density of the Universe and $f(\sigma)$ is an analytical fitting formula. The cumulative number density of haloes above the mass $M_{200c}$ is:
\begin{equation}
n(> M_{200c}) = \int^\infty_{M_{200c}} \frac{{\rm d} n}{{\rm d}\log_{10} M_{200c}} {\rm d}\log_{10} M_{200c}\,. 
\end{equation}
We compare the fitting formula of \citet{Tinker:2010my} (hereafter Tinker10) to the simulation results. Tinker10 calibrated their fitting formula using a SO algorithm to identify dark matter haloes in numerical simulations which is consistent with the approach used in {\sc rockstar}. The analytical predictions were computed by using the online tool {\sc HMFcalc}\footnote{\url{http://hmf.icrar.org/}} \citep{Murray:2013qza}.

\begin{figure}
	\centering
\includegraphics[width=0.49\textwidth]{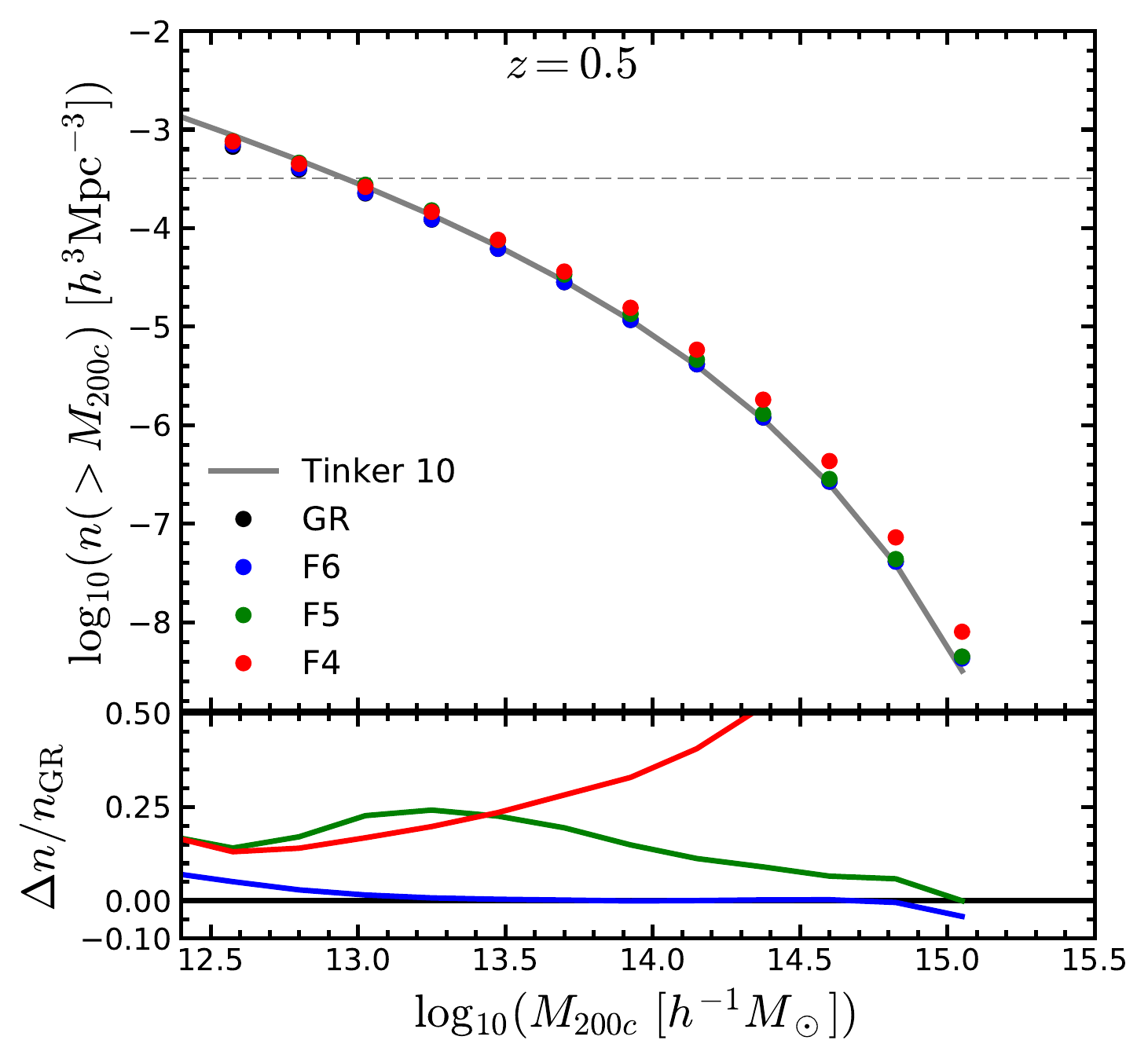}
	\caption{The cumulative halo mass function in the models at $z=0.5$. Different colours represent different models, as labelled. The values for each model correspond to the average over the 5 realisations. The horizontal dashed line shows the number density we use to define our halo sample ($n_{\rm h} = 3.2 \times 10^{-4}~h^3 \, \mathrm{Mpc}^{-3}$). The grey curve shows the Tinker10 cHMF for GR at $z=0.5$. The lower panel shows the relative difference with respect to the $\Lambda$CDM (GR) model.}
	\label{fig:HMF}
\end{figure}

Fig.~\ref{fig:HMF} shows the cumulative halo mass function (cHMF) measured from the simulations and the relative difference between the $f(R)$ models and GR at $z=0.5$.  As expected, the largest deviation from GR is displayed by the F4 model (red line; \citealt{Schmidt:2008tn,Lombriser:2013wta,Cataneo:2016iav}).  The lower panel of Fig.~\ref{fig:HMF} shows that the cumulative halo mass function in the F4 model reaches a difference with respect to GR of $>$50 percent for haloes of mass $M_{200c} > 10^{14.3} h^{-1} {\rm M_{\odot}}$. The maximum difference found between F5 (green line) and GR reaches 25 percent for haloes with mass $M_{200c} \approx 10^{13.2} h^{-1} M_\odot$. On the other hand, for the F6 model (blue line) we see that for very massive haloes, the halo mass function is the same as that in GR. This is because the chameleon mechanism works efficiently for such haloes to suppress the effects of the enhancement to gravity. These differences are purely the result of the modified gravitational force in $f(R)$ models. The stronger deviation of F4 from GR is due to the inefficient screening mechanism in this model compared to the screening in the F6 model.

To make a direct comparison between the halo and galaxy clustering we select a halo population from the simulations by fixing the halo number density (in this case we take the number density of the BOSS-CMASS-DR9 sample at $z=0.5$, $n_{\rm h} = n_{\rm g} = 3.2 \times 10^{-4}~h^3 \mathrm{Mpc}^{-3}$; \citealt{Anderson:2012sa}) and selecting haloes above the mass threshold corresponding to that number density. The horizontal dashed line in Fig.~\ref{fig:HMF} corresponds to the halo number density used to define our halo sample. The minimum mass that defines the halo sample for each model is: $7.643 \times 10^{12} h^{-1}M_\odot$ (GR), $7.798 \times 10^{12} h^{-1} {\rm M_\odot}$ (F6), $9.124 \times 10^{12} h^{-1} {\rm M_\odot}$ (F5) and  $8.734 \times 10^{12} h^{-1} {\rm M_\odot}$ (F4). The fact that F5 has a higher minimum mass is because this model produces more haloes with mass $M_{200c} \sim 10^{13} h^{-1} {\rm M_\odot}$ (as we can see from the lower panel of Fig.~\ref{fig:HMF}) than F6 and F4. For F4, many of the medium-mass haloes have merged to form more massive haloes, hence this model predicts fewer smaller haloes than F5.

\subsection{HOD prescription and galaxy catalogues}\label{sub:hod_c}

\begin{table*}
\centering
\caption{Values of the HOD parameters (columns 2-6) for $f(R)$ models (F6, F5, F4) at $z=0.5$ and  different realisations (Box 1 $-$ Box 5).}
\label{tab:hod}
\begin{tabular}{cccccc} \hline \hline
           & $\log_{10}(M_{\rm min}/[h^{-1} {\rm M_\odot}])$ & $\log_{10}(M_1/[h^{-1} {\rm M_\odot}])$ & $\log_{10}(M_0/[h^{-1} {\rm M_\odot]})$ & $\sigma_{\log M}$ & $\alpha$ \\ \hline
F6 (Box 1) & 13.092 & 14.004 & 13.082 & 0.538 & 1.0125  \\
F6 (Box 2) & 13.093 & 14.004 & 13.081 & 0.539 & 1.0128  \\
F6 (Box 3) & 13.092 & 14.006 & 13.083 & 0.554 & 1.0131  \\
F6 (Box 4) & 13.093 & 14.002 & 13.082 & 0.545 & 1.0132  \\
F6 (Box 5) & 13.094 & 14.008 & 13.078 & 0.547 & 1.0129  \\ \hline
F5 (Box 1) & 13.101 & 14.050 & 13.077 & 0.434 & 1.0643  \\
F5 (Box 2) & 13.100 & 14.045 & 13.078 & 0.449 & 1.0674  \\
F5 (Box 3) & 13.134 & 14.043 & 13.085 & 0.525 & 1.0982  \\
F5 (Box 4) & 13.110 & 14.041 & 13.078 & 0.470 & 1.0710  \\
F5 (Box 5) & 13.108 & 14.041 & 13.080 & 0.462 & 1.0440  \\ \hline
F4 (Box 1) & 13.084 & 14.092 & 13.076 & 0.394 & 1.0921  \\
F4 (Box 2) & 13.075 & 14.113 & 13.073 & 0.345 & 1.0804  \\
F4 (Box 3) & 13.063 & 14.113 & 13.072 & 0.333 & 1.0974  \\
F4 (Box 4) & 13.076 & 14.109 & 13.070 & 0.353 & 1.1110  \\ 
F4 (Box 5) & 13.053 & 14.105 & 13.075 & 0.290 & 1.1143  \\ \hline
\hline 
\end{tabular}
\end{table*} 

To compare the simulations with observations one has to populate dark matter haloes with galaxies. This can be done using one of a number of empirical techniques depending on the physical application we are interested in, such as subhalo abundance matching \citep{Vale:2004yt,Conroy:2005aq,Reddick:2012qy,Klypin:2013rsa}, the conditional luminosity function \citep{Yang:2002ww,Cooray:2005yt} or the halo occupation distribution \citep{Berlind:2001xk,Kravtsov:2003sg,Zheng:2004id}. These empirical descriptions of the galaxy-halo connection have the flexibility to give accurate reproductions of observational estimates of galaxy clustering.   
A second, more expensive but physically motivated method is hydrodynamical simulation \citep{Schaye:2014tpa,Vogelsberger:2014kha}. A third possibility, which retains the physical basis of hydrodynamical simulation at a fraction of the computational cost is semi-analytical modelling of galaxy formation \citep{Somerville:1998bb,Cole:2000ex,Baugh:2006pf,Benson:2010de} in which an N-body dark matter-only simulation is populated with galaxies after solving a set of coupled differential equations. To date, little work has been done to study galaxy formation and clustering in modified gravity models (see  \citealt{Fontanot:2013a} for an example), so here we will resort to the empirical approach of HOD modelling. 

We populate haloes using a functional form for the halo occupation distribution (HOD; \citealt{Peacock:2000qk,Berlind:2001xk}) with five parameters, as used by \citet{Zheng:2007zg}. 

In this form, the mean number of galaxies in a halo of mass $M_{\rm h}$ (in our case $M_{\rm h} = M_{200c}$) is the sum of the mean number of central galaxies plus the mean number of satellite galaxies,
\begin{eqnarray}\label{eq:Ng}
\left\langle N(M_{\rm h}) \right\rangle &=& \left\langle N_{\rm c}(M_{\rm h}) \right\rangle + \left\langle N_{\rm s}(M_{\rm h}) \right\rangle\,,\\
\left\langle N_{\rm c}(M_{\rm h}) \right\rangle &=& \frac{1}{2} \left[ 1 + \mathrm{erf}\left( \frac{\log_{10} M_{\rm h} - \log_{10} M_\mathrm{min}}{\sigma_{\log M}} \right) \right]\,,\label{eq:Nc}\\
\left\langle N_{\rm s}(M_{\rm h}) \right\rangle &=& \left\langle N_{\rm c}(M_{\rm h}) \right\rangle \left( \frac{M_{\rm h} - M_0}{M_1} \right)^{\alpha}\,,\label{eq:Ns}
\end{eqnarray}
and $\left\langle N_{\rm s}(M_{\rm h}) \right\rangle = 0$ if $M_{\rm h} < M_0$. $\left\langle N_{\rm c/s}(M_{\rm h}) \right\rangle$ is the average number of central or satellite galaxies, respectively, in a halo of mass $M_{\rm h}$. The model depends on five parameters: $M_{\rm min}$, $M_0$, $M_1$, $\sigma_{\mathrm{log} M}$ and $\alpha$. From Eqns.~\eqref{eq:Nc} and \eqref{eq:Ns} we can see that $M_\mathrm{min}$ and $M_0$ represent the halo mass threshold to host one central or one satellite galaxy, respectively. Also, we assume that central galaxies are placed at the centre of their host haloes and satellite galaxies are orbiting inside haloes with $M_{\rm h} \geq M_0$. The satellite galaxies are radially distributed, between $r=[0,r_{200c}]$, following the Navarro-Frenk-White (NFW) profiles of their host halo \citep{Navarro:1995iw,Navarro:1996gj}.

We generate five galaxy catalogues (one for each independent realisation of the density field) for every gravity model following the prescription described above. The galaxy catalogues match the galaxy number density of the BOSS-CMASS-DR9 sample at $z=0.5$ ($n_{\rm g} = 3.2 \times 10^{-4}~h^3 \, \mathrm{Mpc}^{-3}$ ; \citealt{Anderson:2012sa}) and the galaxy two-point correlation function across all gravity models (more details are presented below). The BOSS-CMASS sample is dominated by LRGs which are massive galaxies typically residing in haloes with $M_h \sim 10^{13} h^{-1}M_\odot$ \citep{Anderson:2012sa}. Hence, given the mass resolution of the {\sc elephant} simulations, these runs are suitable to study the impact of $f(R)$ gravity on galaxy clustering. 

The values of the HOD parameters used to populate the GR simulations are those inferred from the abundance and clustering measured for the BOSS-CMASS-DR9 galaxy sample \citep{Manera:2012sc}:
\begin{eqnarray}\label{eq:HOD_v}
\log_{10}(M_{\rm min}/[h^{-1}M_\odot]) &=& 13.09\,,\nonumber\\
\log_{10}(M_1/[h^{-1}M_\odot]) &=& 14.00\,,\nonumber\\
\log_{10}(M_0/[h^{-1}M_\odot]) &=& 13.077\,,\nonumber\\
\sigma_{\log M} &=& 0.596\,,\nonumber\\
\alpha &=& 1.0127.
\end{eqnarray}

To find the $f(R)$ HOD parameters, we use the simplex algorithm of \citet*{Nelder:1965} to search through the 5D parameter space. We start the algorithm with an initial guess at the values of  the HOD parameters, then the code walks through the 5-dimensional HOD parameter space looking for the values that minimize the root-mean-square difference ($\it rms$) of the two-point correlation function between $f(R)$ and GR models. We measure the correlation function using 40 logarithmically spaced radial bins between $1 - 80~h^{-1}{\rm Mpc}$. The fractional difference of the galaxy number density is also included (with a weight of 8) in the ${\it rms}$ value calculated in order to ensure similar numbers of galaxies in all catalogues. We stop the search when ${\it rms} < 0.02$ (this means that the overall agreement is better than $2\%$). For the F4 model, the minimum value of the {\it rms} we could obtain in practice was $\sim 0.03$.

\begin{figure*}
	\centering
\includegraphics[width=0.49\textwidth]{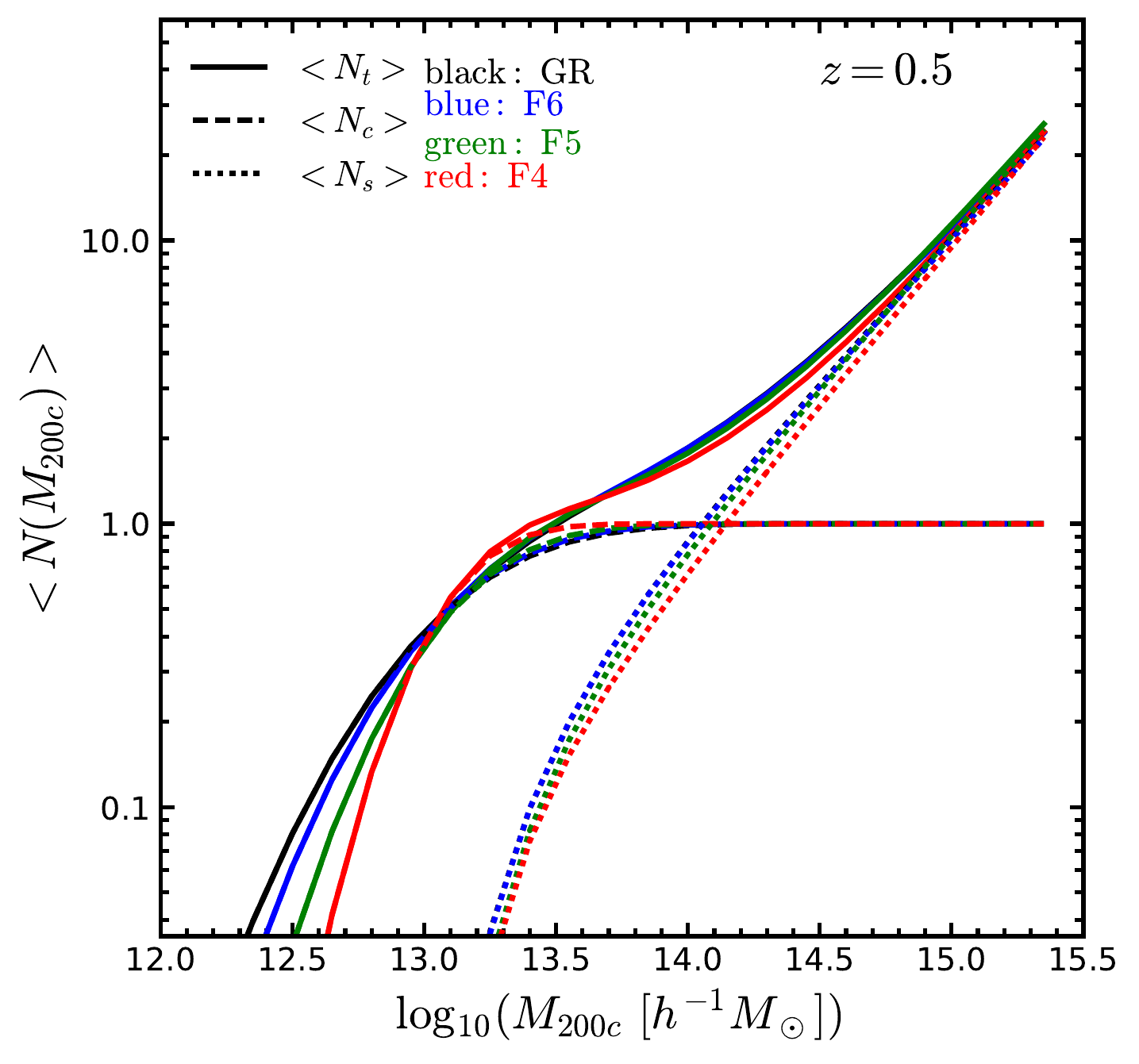}
\includegraphics[width=0.5\textwidth]{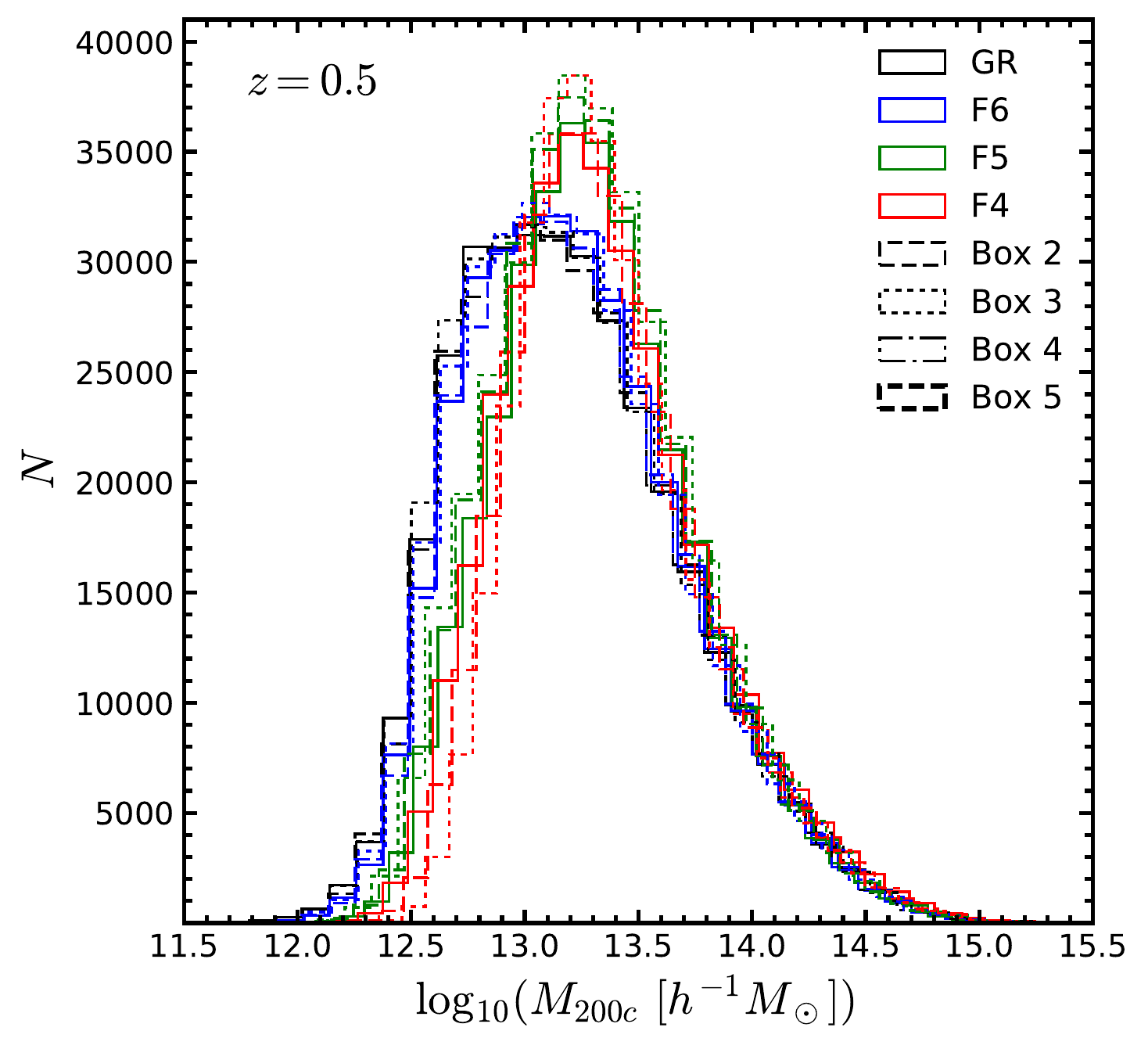}
    \caption{Left panel: The mean number of central and satellite galaxies as a function of halo mass,  $\left\langle N_{c/s}(M_{200c}) \right\rangle$. Dashed lines show the HOD for central galaxies and dotted lines show satellite galaxies while solid lines represent the total averaged number of galaxies, calculated from Eqs.~\eqref{eq:Ng} -- \eqref{eq:Ns} with parameters \eqref{eq:HOD_v} for GR and the result of Box 1 listed in Table~\ref{tab:hod} for the $f(R)$ models, as labelled. Right panel: the number of galaxies in the simulation as a function of the host halo mass, the same distribution at $z=0.5$ for different realisations: Box 1 (solid lines), Box 2 (dashed lines), Box 3 (dotted lines), Box 4 (dashed-dotted lines) and Box 5 (thick-dashed lines).}
    \label{fig:HOD}
\end{figure*}

Here, we are interested in the marked correlation function (mCF), which was proposed to highlight the environmental dependence in modified gravity models \citep{White:2016yhs}. Hence, the most natural choice is to make the unmarked two-point correlation functions (2PCFs) of the different models as close to each other as possible; otherwise when there is a difference in their mCFs we can not be sure how much of this is due to the different 2PCFs. 

The $f(R)$ HOD parameters were tuned for each model and realisation to match the clustering displayed in the counterpart simulation from the GR suite. The best-fitting values of the HOD parameters for the different realisations are listed in Table~\ref{tab:hod}. The variation in the best-fitting parameter values is larger as the modification to gravity increases. We note that, despite the differences between the values of the best-fitting parameters between different models and realisations, the resulting correlation functions and galaxy number densities agree with the GR results to within our target accuracy. 

Note that the HOD parameters are degenerate to some extent, so that a comparison of the values of any single parameter across realisations or models should not be over interpreted. For instance, consider the parameter $\alpha$ that governs the number of satellite galaxies in haloes of a given mass: in the case of the most extreme model, F4, the variation between realisations is $\sim 3\%$. This difference is small compared with the $1\sigma$ scatter of HOD parameter fittings, e.g., \citet{White:2010ed}.   

The left panel of Fig.~\ref{fig:HOD} shows the HOD for CMASS galaxies at $z=0.5$. The gradual transition from zero to one galaxy per halo is determined by the values adopted for $\log M_{\rm min}$ and $\sigma_{\log M}$ for central galaxies (dashed lines). The appearance of satellites in haloes (dotted lines) is dictated by the values of $M_1$ and $M_0$, and the rapid increase in the satellite content of haloes with increasing halo mass is governed by $\alpha$. The HOD parameters are adjusted in the $f(R)$ models to approximately reproduce the abundance and clustering of CMASS galaxies realised in GR. We note that the resulting HODs are very similar between $f(R)$ gravity and GR.

The right panel of Fig.~\ref{fig:HOD} shows the distribution of the number of galaxies as a function of the host halo mass $(M_{200c})$. We see that most galaxies are found in haloes with mass $10^{13} < M_{200c}/[h^{-1} {\rm M_\odot}] < 10^{14}$. We also note that the F5 and F4 models produce more galaxies than GR and F6 in this mass range. This is because the abundance of haloes in this mass range is boosted in F5 and F4, as we can see from the relative differences of the cHMFs presented in the lower panel of Fig.~\ref{fig:HMF}. Analysing the distribution of galaxies, we find good agreement between the five realisations. 

From the distribution of galaxies as a function of host halo mass plotted in Fig.~\ref{fig:HOD} we note that $\approx 0.1\%$ galaxies reside in poorly resolved haloes $(M_{200c} < 10^{12}h^{-1} M_\odot)$. The inclusion of these galaxies in the final catalogues does not affect the clustering results.

\begin{figure*}
	\centering
\includegraphics[width=0.49\textwidth]{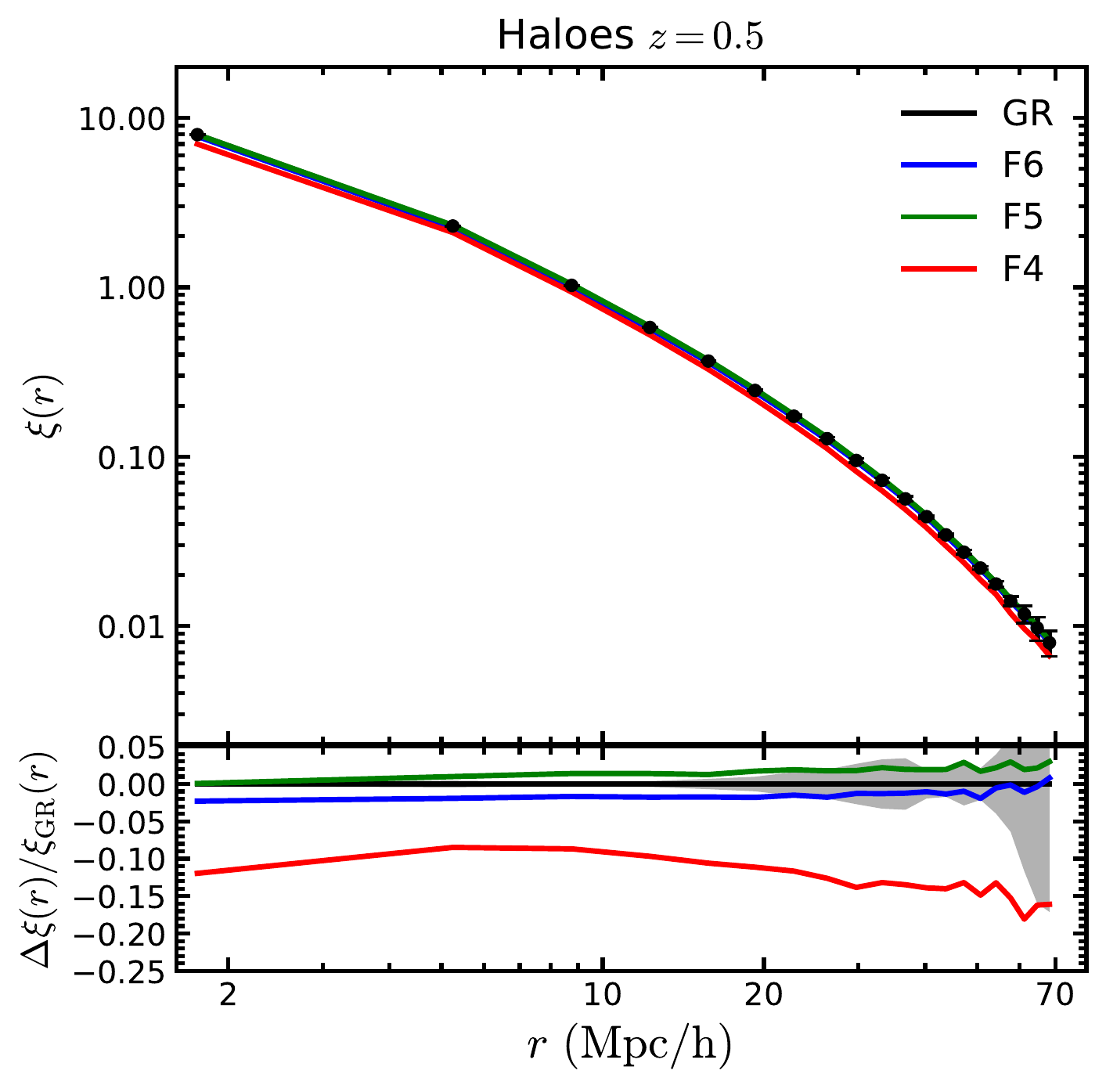}
\includegraphics[width=0.5\textwidth]{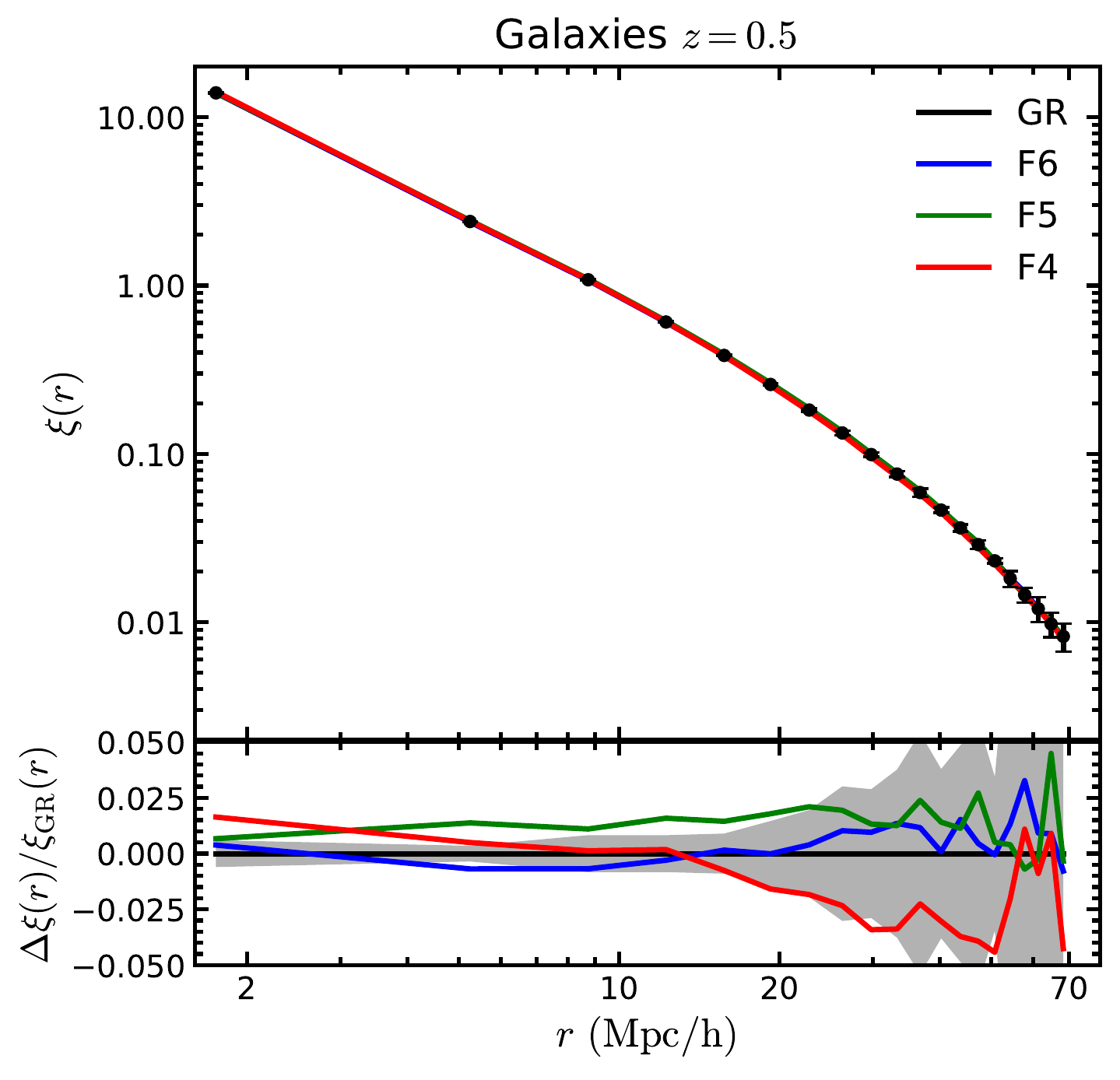}
	\caption{Two--point correlation function in real space measured for haloes (left panel) with $n_{\rm h} = 3.2\times 10^{-4} h^3\mathrm{Mpc}^{-3}$ and HOD galaxies (right panel) in the four gravity models at $z=0.5$. The plotted values correspond to the average of the 5 realisations for each model. Different colour lines correspond to different gravity models as labelled.
    The lower subpanels show the relative difference between the results from the $f(R)$ and $\Lambda$CDM (GR) models. Error bars and shaded regions correspond to $1\sigma$ standard deviation over the 5 GR realisations.}
	\label{fig:2pcf}
\end{figure*}
\section{Results}\label{sec:halo}
Upcoming galaxy surveys will allow us to measure the clustering of galaxies to an unprecedented level of accuracy with the aim of developing a better understanding of the nature of dark matter, dark energy and the evolution of galaxies through cosmic time. 
In this section we present the statistical tools which can be used to characterise the halo and galaxy distributions in different gravity models. 
This is the first time that the halo and galaxy clustering has been studied to this level of detail in $f(R)$ gravity models. 

\subsection{$2$-point correlation function}\label{sub:2pcf}
To characterise the clustering of dark matter haloes and galaxies, we use the two-point correlation function, $\xi(r)$. This is defined as the excess probability, compared with that expected for a random distribution, of finding two haloes (or galaxies) contained in volume elements $\mathrm{d}V_1$ and $\mathrm{d}V_2$ at a separation $r$ 
\citep{Peebles80}:

\begin{equation}\label{eq:Pxi}
\mathrm{d}P_{12}(r) \equiv \bar{n}^2 [1 + \xi(r)]\mathrm{d}V_1 \mathrm{d}V_2\,,
\end{equation}
where $\bar{n}$ is the mean halo (galaxy) number density. The 2-point correlation functions and therefore the marked correlation functions are measured within the range $1-80\,h^{-1}$Mpc (for details see Sec.~\ref{sub:hod_c}).

First we study the clustering of dark matter haloes, $\xi_{\rm h}$, (left panel of Fig.~\ref{fig:2pcf}), for our halo samples with $n_{\rm h} = 3.2\times 10^{-4} h^3\mathrm{Mpc}^{-3}$. 
Although this statistic is not directly observable, it is instructive to study the properties of $\xi_{\rm h}$, since this is a first step towards understanding differences in the clustering of galaxies.  

The first thing we notice is that the deviation from GR does not show a monotonic dependence on $|f_{R0}|$. More explicitly, F6 and F4 models have weaker clustering than GR, while F5 haloes are more clustered than GR. 

These perhaps counterintuitive results can be explained by considering the following two effects of the enhanced gravity.

Firstly,  stronger gravity means a faster growth of initial density peaks, and therefore more massive structures at late times. This generally leads to a higher mean halo number density above a fixed halo mass threshold. The enhancement of halo formation is not uniform: when screening is efficient, it is stronger in low-density regions than it is in high-density regions; when the screening is less efficient, then the growth of haloes is boosted in all environments, and those in dense regions can be boosted more because they have more matter around them to accrete.

Secondly, enhanced gravity generally leads to a stronger clustering of the structures that are formed from these initial density peaks. However, stronger gravity also means that we can expect more mergers in dense regions, reducing the number of haloes there. The latter effect can be seen by comparing the cHMFs of F5 and F4 in Fig.~\ref{fig:HMF}.

As we choose the halo mass cut to ensure that we consider the same number of haloes in each model any  differences in $\xi_{\rm h}$ come from the different spatial distributions of haloes in the  models. For F6, the deviation from GR is weak and the fifth force is suppressed in high-density regions. As a result small density peaks in low-density regions grow faster than similar density peaks in high-density regions, and more of them make it into the fixed number density halo catalogue than in GR. This makes the haloes less clustered and $\xi_{\rm h}(r)$ smaller.

For F5, the enhancement of gravity is stronger and the screening is weaker, so that haloes in all regions experience faster growth; those in high-density environments have a larger supply of raw materials for accretion and growth, so that they are more likely ending up in the final halo catalogue, leading to a stronger clustering and $\xi_{\rm h}$. For F4, the even stronger enhancement of gravity causes more mergers of haloes in dense regions to form even larger haloes, and to maintain the same $\bar{n}_{\rm h}$ more haloes in low-density regions have to be included into the halo catalogue, leading to less clustering and smaller $\xi_{\rm h}$.

In the case of the galaxy correlation function (right panel  of Fig. \ref{fig:2pcf}), as we said before, the HOD catalogues for the $f(R)$ models were created by tuning the parameters (\ref{eq:HOD_v}; see Table~\ref{tab:hod}) to approximately match the two-point correlation function in GR (to within $1-3\%$).

\begin{figure*}
 \centering
\includegraphics[width=0.32\textwidth]{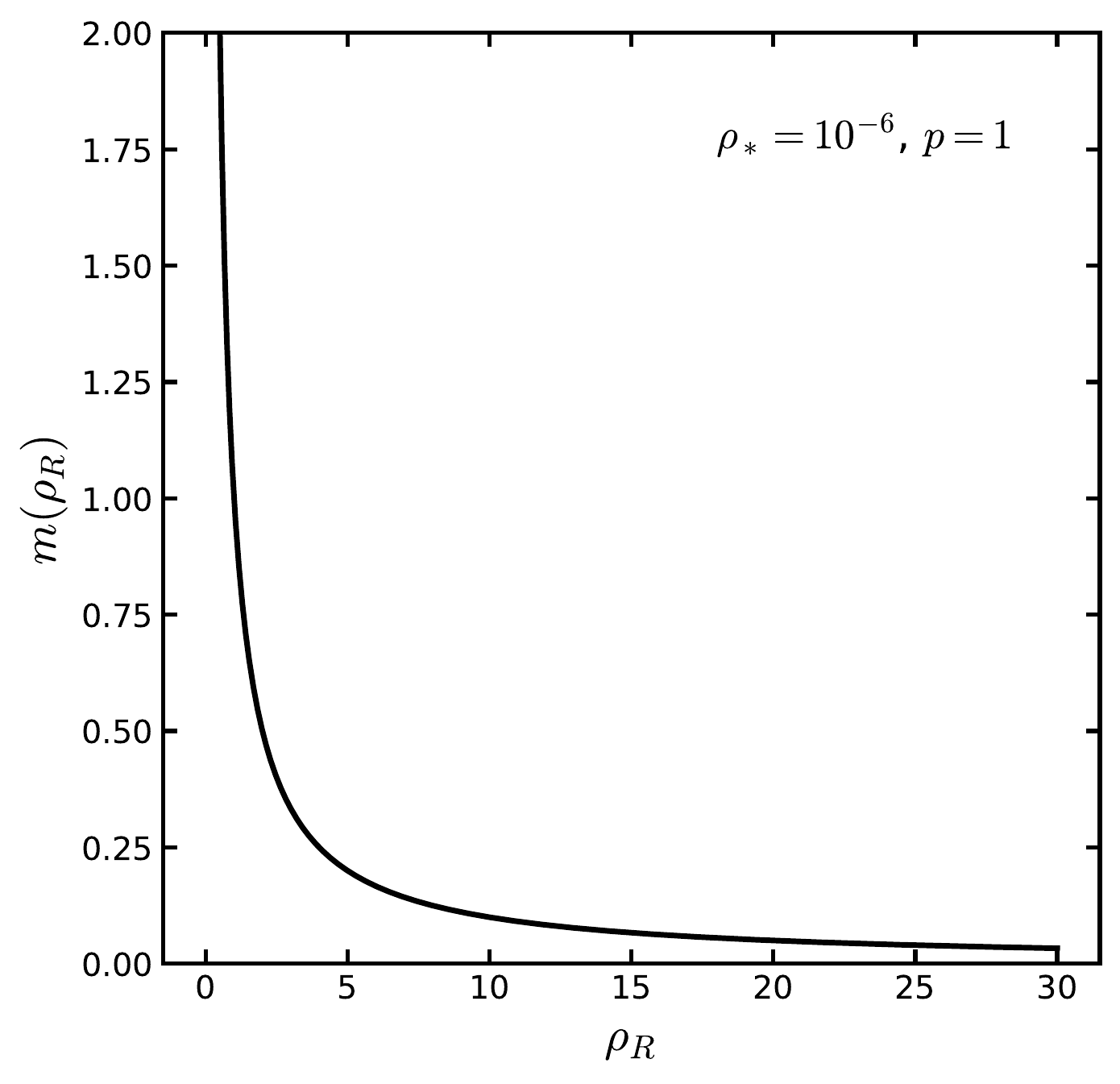} 
\includegraphics[width=0.33\textwidth]{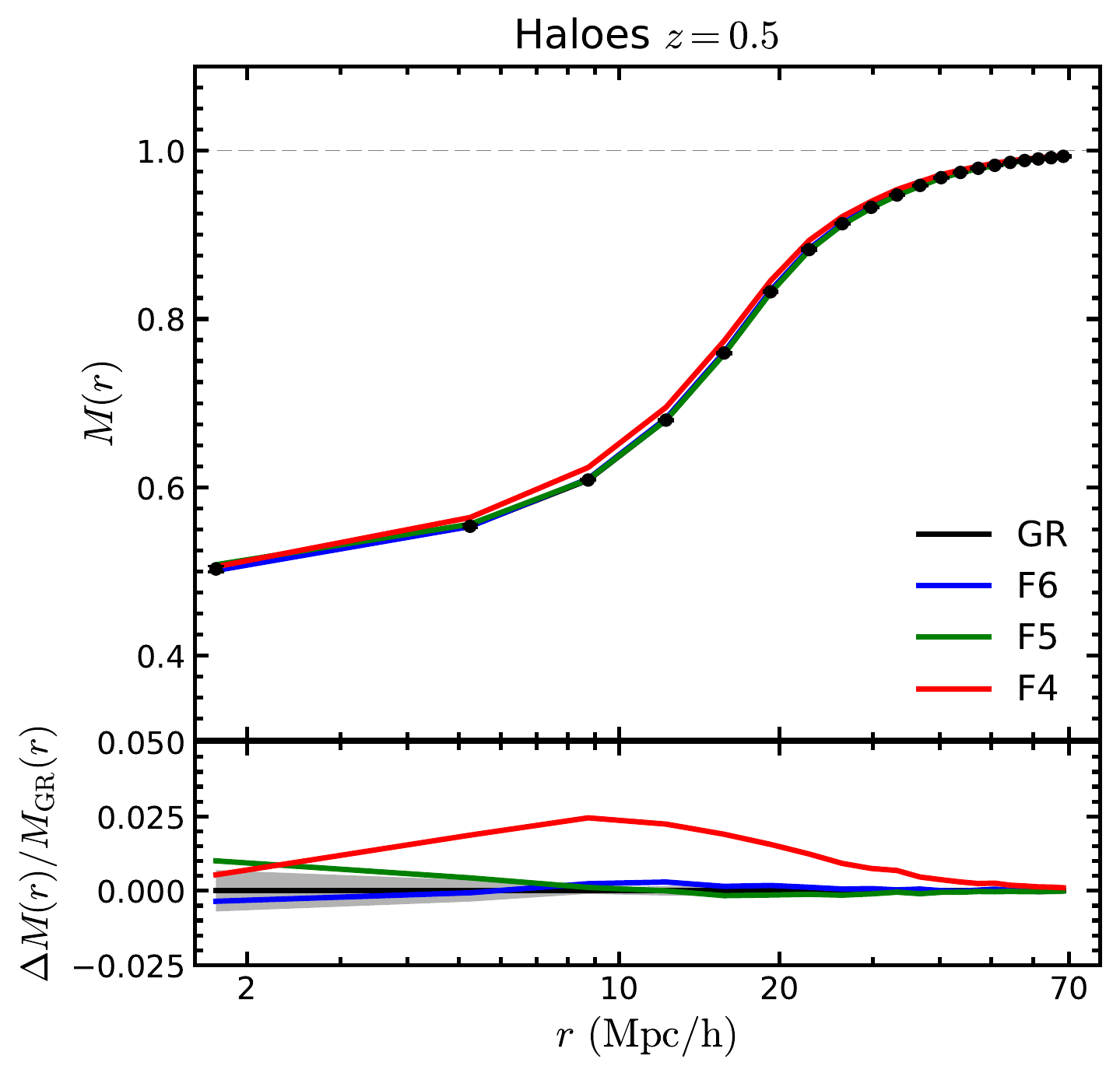}
\includegraphics[width=0.33\textwidth]{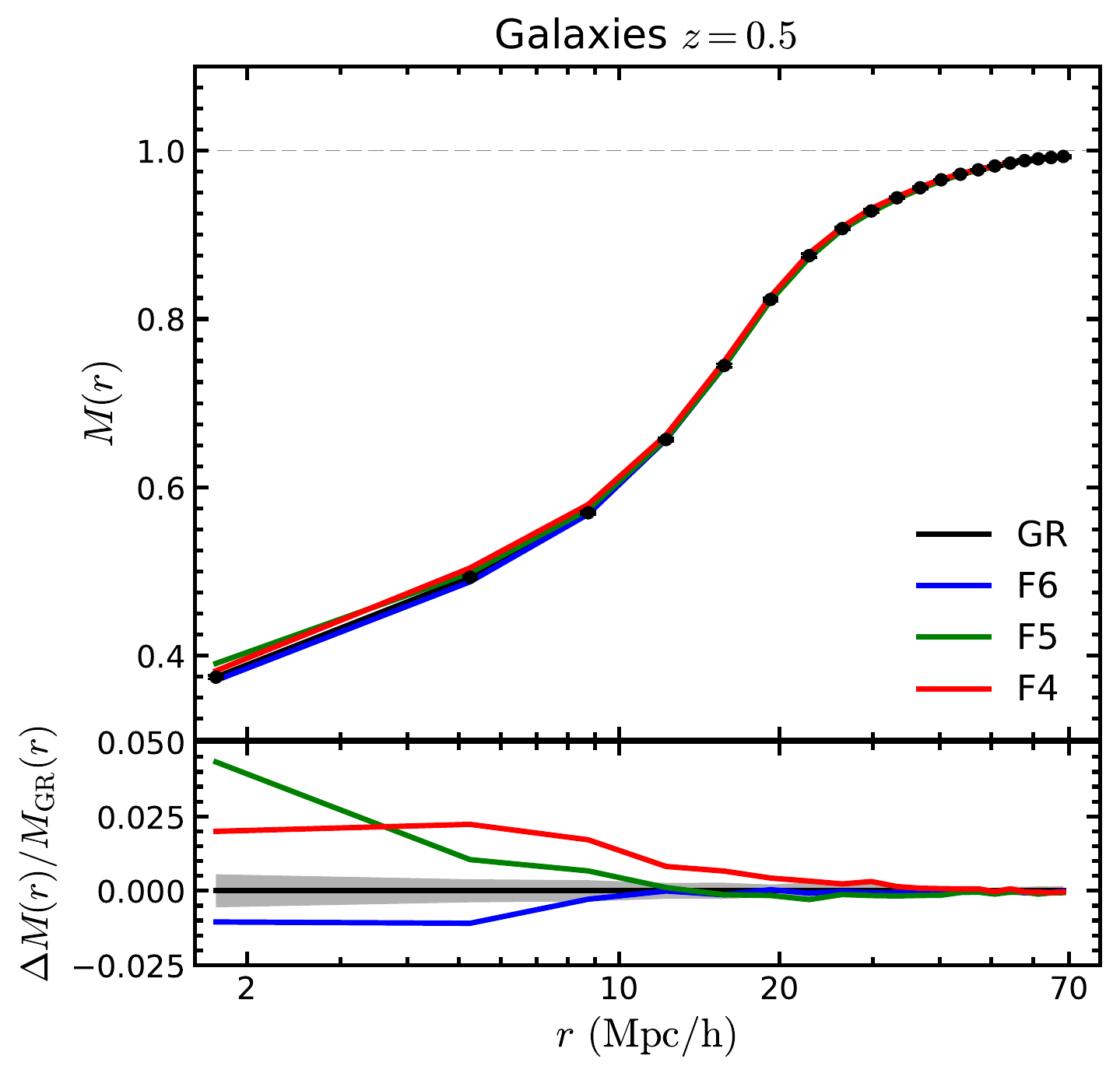}
\includegraphics[width=0.32\textwidth]{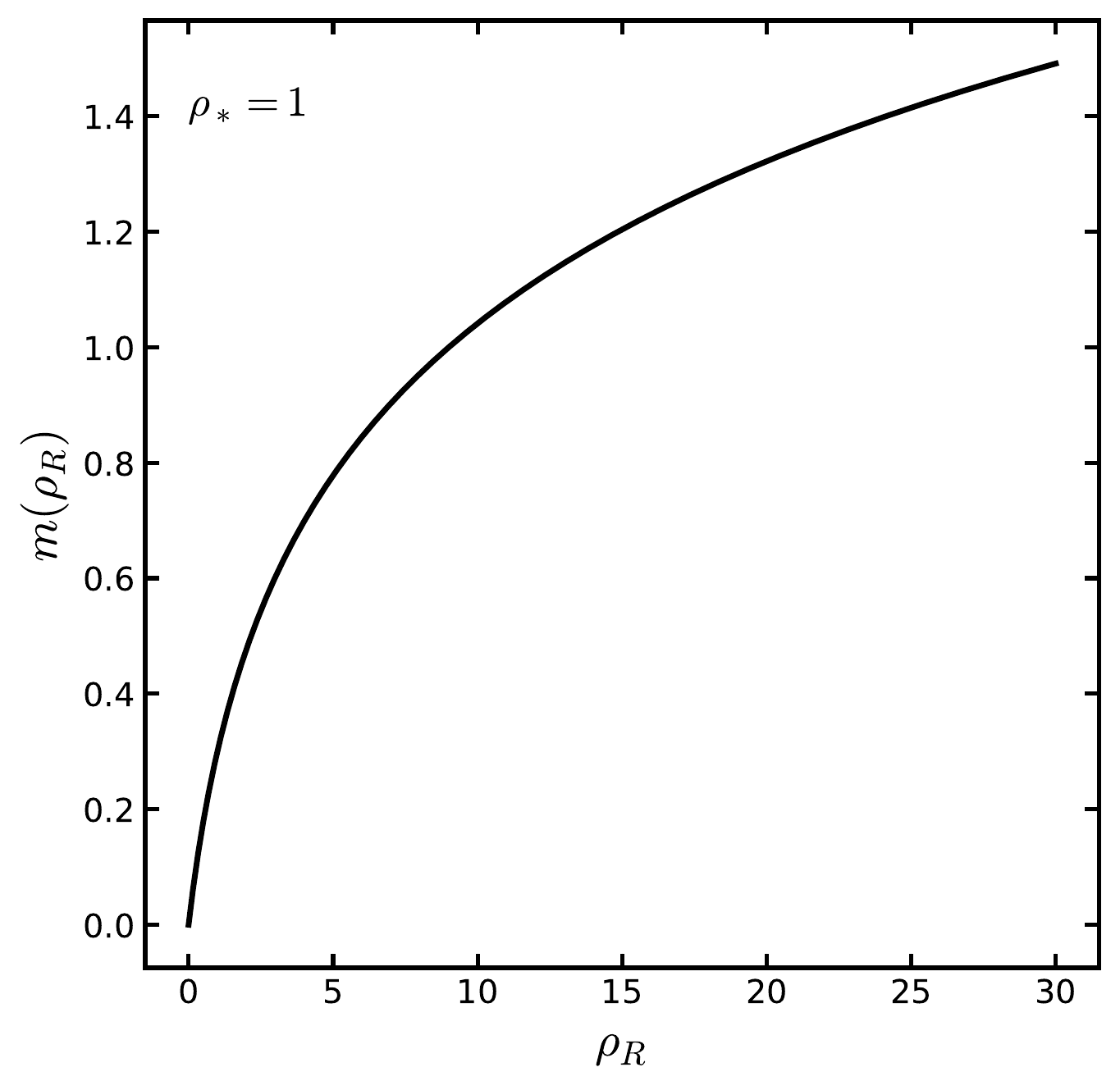} 
\includegraphics[width=0.33\textwidth]{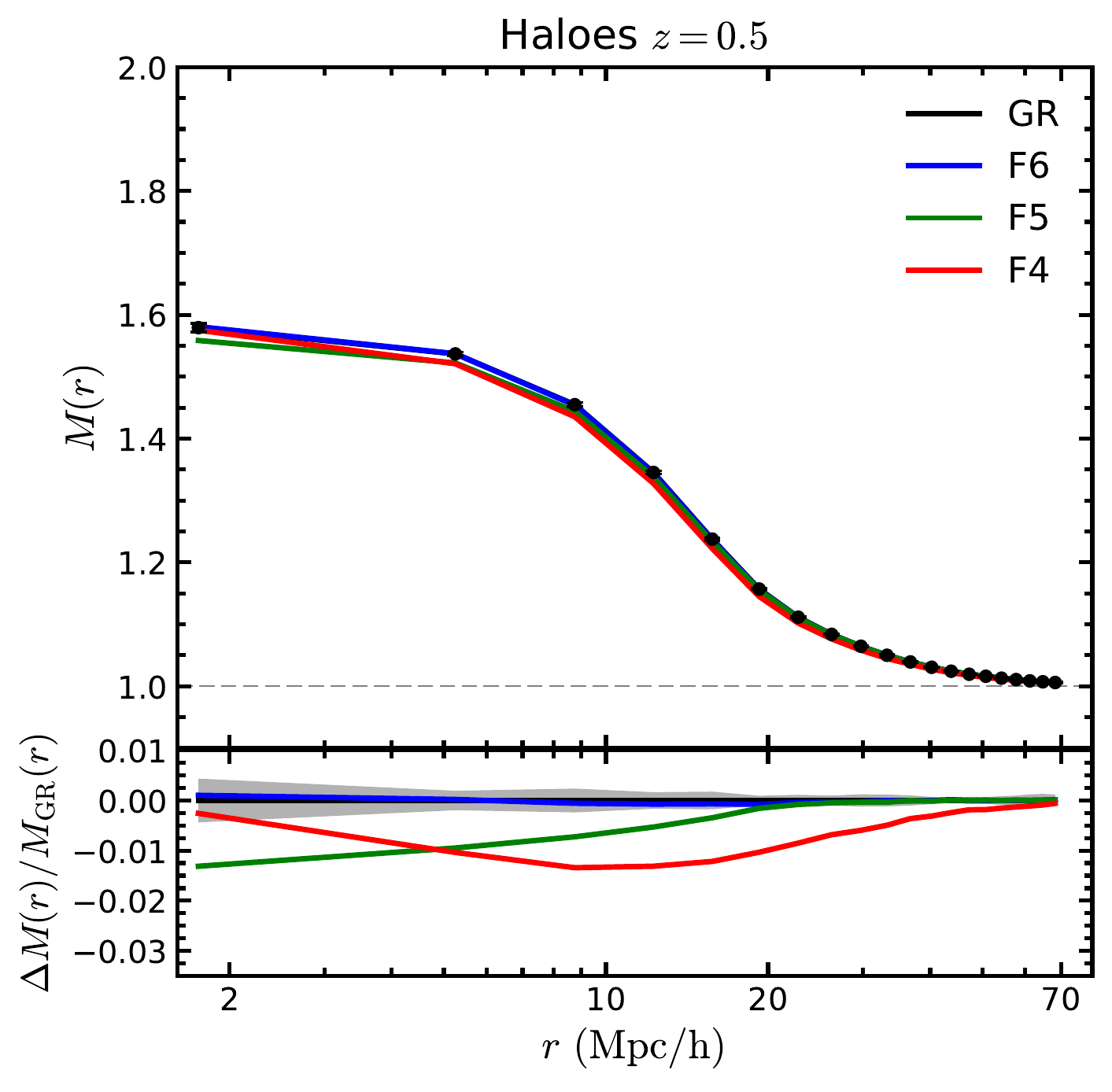}
\includegraphics[width=0.33\textwidth]{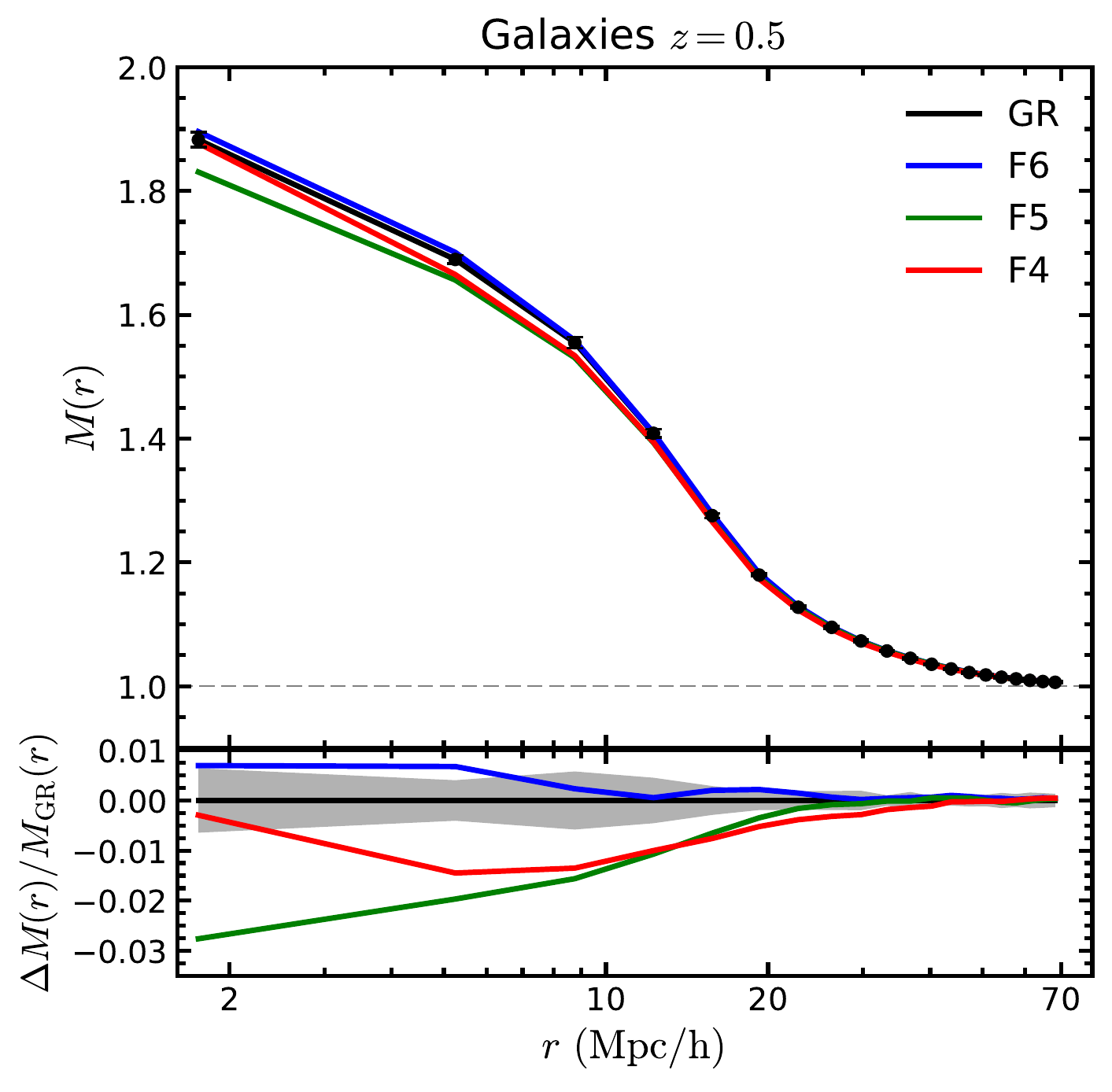}
\includegraphics[width=0.33\textwidth]{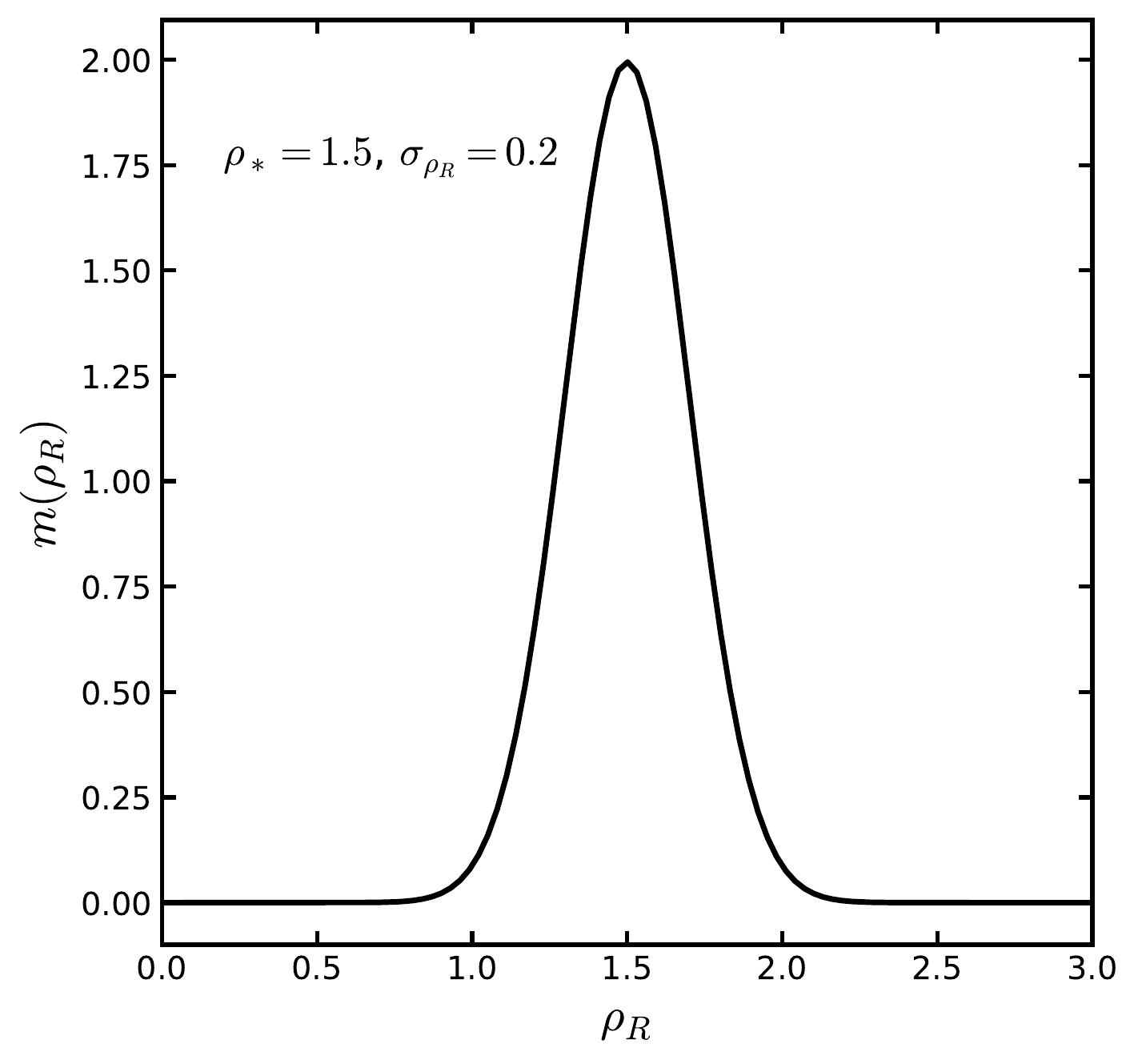}
\includegraphics[width=0.33\textwidth]{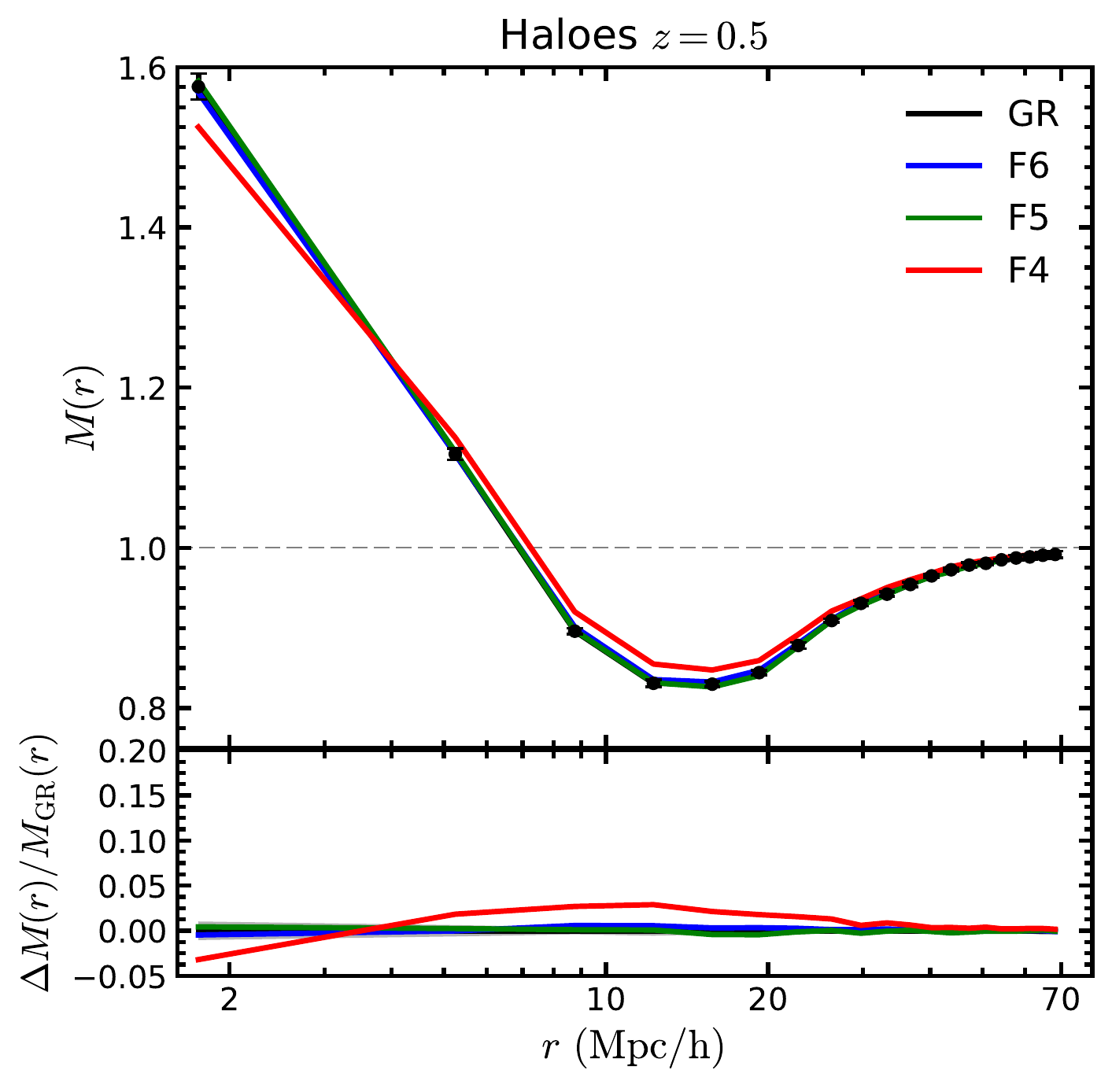}
\includegraphics[width=0.33\textwidth]{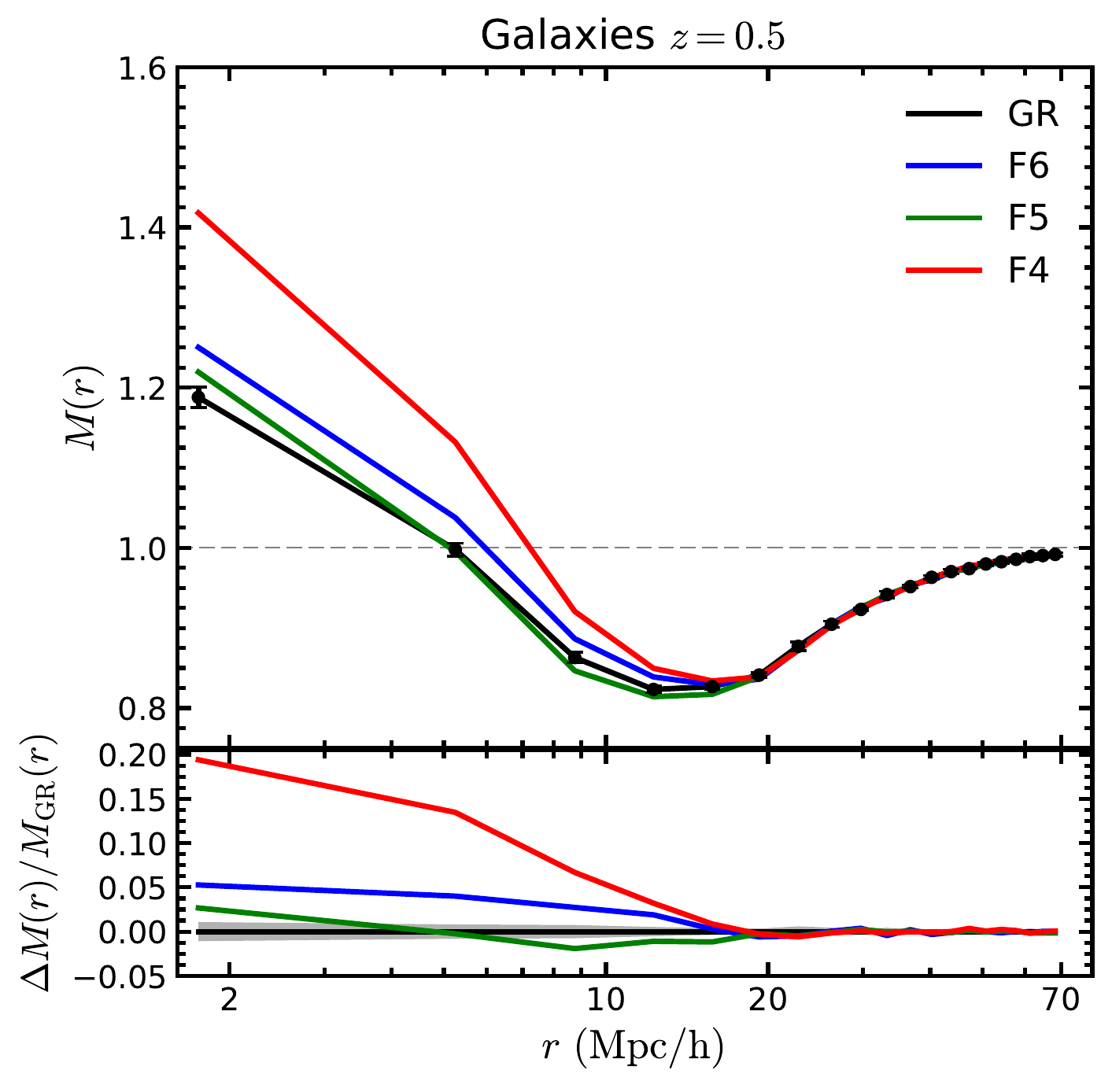}
	\caption{Marked correlation functions of haloes and CMASS galaxies at $z=0.5$; mark in function of number density field. Left: functional form of the mark in function of density field, middle: halo marked correlation functions and right: galaxy marked correlation functions. Plots from upper to bottom: White mark \eqref{eq:W16}, log--mark \eqref{eq:log} and Gaussian-$\rho_R$ mark \eqref{eq:gauss}. All lower subpanels show the relative difference between $f(R)$ models and GR. The plotted values correspond to the average over the 5 realisations. Errors correspond to $1\sigma$ standard deviation over the 5 GR realisations.}
	\label{fig:mCF_rho}
\end{figure*}

\subsection{Marked correlation function}\label{sub:mcf}
In this subsection we consider the marked correlations, in which one weights galaxies\footnote{For simplicity, we talk about galaxies here, but the same calculation can (and will) be applied to haloes.} by some property or `mark' when estimating clustering statistics. Marked correlations are particularly well-suited to quantifying how the properties of galaxies correlate with environment \citep{Sheth:2005aj,Skibba:2005kb,White:2008ii,Skibba:2008fy,Skibba:2011ax}. 
Here, we test the idea proposed by \citet{White:2016yhs} that marked correlation functions may show a  clearer signature of modified gravity in the large-scale clustering of galaxies, by up-weighting low density regions, where screening is weak and deviations from GR are strong.   

The marked correlation function is defined as \citep{Sheth:2005aj}:
\begin{equation}\label{eq:Mr}
M(r) \equiv \frac{1}{n(r)\bar{m}^2} \sum_{ij} m_i m_j = \frac{1 + W(r)}{1 + \xi(r)}\,,
\end{equation}
where the sum is over all pairs with a given separation, $r$, $n(r)$ is the number of such pairs and $\bar{m}$ is the mean mark for the entire sample. In the second equality $\xi(r)$ is the two-point correlation function in which all galaxies (or haloes) are weighted equally. $W(r)$ is derived from a similar sum over galaxy (halo) pairs separated by $r$, as used to estimate $\xi(r)$, but now each member of the pair is weighted by the ratio of its mark to the mean mark of the full sample. 
The marked correlation function $M(r)$ can be estimated approximately using the simple pair count ratio ${\rm WW}/{\rm DD}$ (where ${\rm DD}$ is the count of data--data pairs and ${\rm WW}$ represents the corresponding weighted counts). Hence, no random catalogue is needed for its computation. 
 
The choice of the mark is flexible and depends on the application. Since we are interested in isolating the effects of the chameleon screening mechanism on structure formation, we study the clustering of HOD galaxies using two different definitions of environment: a) the number density field and b) the Newtonian gravitational potential \citep{Shi:2017pyd}.
\subsubsection{Density}\label{sub:den}
For this environment definition we use three marks with an adjustable dependence on density:
\begin{eqnarray}
m &=& \left( \frac{\rho_* + 1}{\rho_* + \rho_R} \right)^p\,,\label{eq:W16}\\
m &=& \log_{10} (\rho_R + \rho_*)\,,\label{eq:log}\\
m &=& \frac{1}{\sqrt{2\pi}\sigma_{\rho_R}} \exp\left(- \frac{(\rho_R - \rho_*)^2}{2\sigma_{\rho_R}^2} \right)\,,\label{eq:gauss} 
\end{eqnarray}
where $\rho_R$ is the galaxy number density in units of the mean galaxy number density, $\bar{\rho}$, and $p$, $\rho_*$ and $\sigma_{\rho_R}$ are adjustable parameters. 

A crucial step in the estimation of the marked correlation function is the definition of the density. 
We measure the galaxy number density using counts-in-cells (see, for example, \citealt{Baugh:1995}). We divide the simulation box into cells (or cubical boxes) of the same size, and then count the number of galaxies inside each cell. Hence, we can compute the overdensity, $\delta$, as:
\begin{equation}\label{eq:rhoR}
1 + \delta = \frac{N}{\bar{N}} \equiv \rho_R\,, 
\end{equation}
where $N$ is the number of galaxies in each cell and $\bar{N}$ is the mean number of galaxies in cells of a given size over the simulation volume. To compute the density we have used $60^3$ cells of size $\sim 17 h^{-1}$ Mpc. Given the mean galaxy number density of CMASS galaxies, $n_{\rm g} = 3.2 \times 10^{-4}~h^3 \mathrm{Mpc}^{-3}$, we have a mean number of galaxies in the cells of $\bar{N} = 1.59$. We checked that reducing the number of cells to $30^3 - 40^3$ does not affect our results significantly, while further reducing the number of cells makes the signal weaker; in the limit of $1^3$ cell, $W(r)$ becomes identical to $\xi(r)$, as expected. 

The first mark, Eqn.~\eqref{eq:W16}, was proposed by \citet{White:2016yhs} (hereafter the White--mark), with the motivation being that by up-weighting low density regions (i.e. by choosing $p > 0$), one might be able to find a signature of modified gravity, since previous studies have shown that the properties of voids are different in modified gravity theories than in GR \citep{Clampitt:2012ub,Cai:2014fma,Zivick:2014uva,Cautun:2017tkc}. 
The log mark, Eqn.~\eqref{eq:log}, allows us to up-weight regions with $\rho_R > 1$, i.e., intermediate and high-density regions. Finally, using the Gaussian-$\rho_R$ mark, Eqn. \eqref{eq:gauss}, we are able to control the regions we want to up-weight. Previously, \citet*{Llinares:2017ykn} found that by using a Gaussian transformation of the density field is it possible to up-weight intermediate density regions and find bigger differences between the clustering of objects in modified gravity and GR models. Keeping this in mind, we use the Gaussian-$\rho_R$ mark to up-weight only intermediate density regions.

It is evident that by using Eqn.~\eqref{eq:W16} one can control the up-weighting by varying the power $p$ and the parameter $\rho_*$. For simplicity we chose $p=1$ and $\rho_* = 10^{-6}$. With the log-mark, a natural choice of the parameter which controls the enhancement is $\rho_* = 1$, given $m=0$ for voids $(\rho_R = 0)$. The parameters we chose for the Gaussian-$\rho_R$ mark are: $\rho_* = 1.5$ and $\sigma_{\rho_R} = 0.2$, which ensures that we up-weight  intermediate-density regions of interest. The functional form of the marks, Eqs. \eqref{eq:W16} -- \eqref{eq:gauss}, is shown in the left-hand panels of Fig.~\ref{fig:mCF_rho}. We have tried using different values of $p$, $\rho_R$ and $\rho_*$ but  found that our results do not show significant differences on varying these parameters. We refer to low-, intermediate- and high-density regions as those for which the cells contain $N = 1$, $2-3$ and $>4$ objects or, equivalently, to cells with $\rho_R = 0.62$, $1.25-1.88$ and $>2.51$, respectively (see Eqn.~\eqref{eq:rhoR}).

Fig.~\ref{fig:mCF_rho} shows the marked correlation functions (mCFs) at $z=0.5$ measured from the halo (middle panels) and the HOD (left panels) catalogues in the $f(R)$ and GR models. In all cases the marked correlation function goes to unity on large scales as expected (see right-hand expression of Eqn.~\eqref{eq:Mr}). The first row of plots in Fig.~\ref{fig:mCF_rho} shows the mCF using the mark defined by Eqn.~\eqref{eq:W16}, the White-mark, with $p=1$ and $\rho_* = 10^{-6}$, the second row shows the log-mark, Eqn.~\eqref{eq:log} with $\rho_*=1$, and the third row shows the Gaussian-$\rho_R$ mark with $\rho_* = 1.5$ and $\sigma_{\rho_R} = 0.2$.  We observe different behaviours: for the White-mark,  Eqn.~\eqref{eq:W16}, the marked correlation function is $M(r) \leq 1$ at small separations, for the log mark, Eqn.~\eqref{eq:log}, we have $M(r) \geq 1$, while for the Gaussian-$\rho_R$ we notice a transition from $M(r) \leq 1$ to $M(r) > 1$ at intermediate scales. 

Analysing the behaviour of the halo marked correlation functions (see middle panel of Fig.~\ref{fig:mCF_rho}) we find the following features: 
\begin{itemize}
\item The clustering of F6 is almost indistinguishable from that of GR for all three marks, because of the efficient screening.

\item For F5, the stronger growth (see Sec.~\ref{sub:2pcf}) means more clustering of haloes on small scales, which is why $W(r)$ and therefore the marked correlation function is more affected at smaller $r$.

\item In the case of F4, the higher production rate of massive haloes, driven by the more frequent mergers of lower mass haloes (see Sec.~\ref{sub:2pcf}), leads to the incorporation of haloes into the fixed number density sample which correspond to low density peaks and which are more likely to come from low-density regions. Hence, the probability of finding a pair of tracers (haloes or galaxies) increases at intermediate separation $r$ due to presence of these low mass haloes.  
\end{itemize}

The right columns of Fig.~\ref{fig:mCF_rho} show that galaxies qualitatively mimic the marked clustering of haloes (at least for the White and log marks). Hence, the behaviour of the galaxy marked correlation functions can be understood following the same explanation as presented above for haloes. It is interesting to notice that even with the added complexity of populating haloes with HOD galaxies, the qualitative behaviour of the marked correlation functions preserves, suggesting that a true physical feature is being observed here.

For the Gaussian$-\rho_R$ mark, Eqn.~\eqref{eq:gauss}, 
which enhances intermediate-density regions (cells with 2 or 3 haloes/galaxies), we found that the F4 galaxy marked correlation function reaches a maximum of $20\%$ for the lowest separation bin, while F6 predicts a difference of $5\%$ and F5 keeps closer to GR with a difference of $\sim 3\%$.

\begin{figure*}
 \centering
\includegraphics[width=0.33\textwidth]{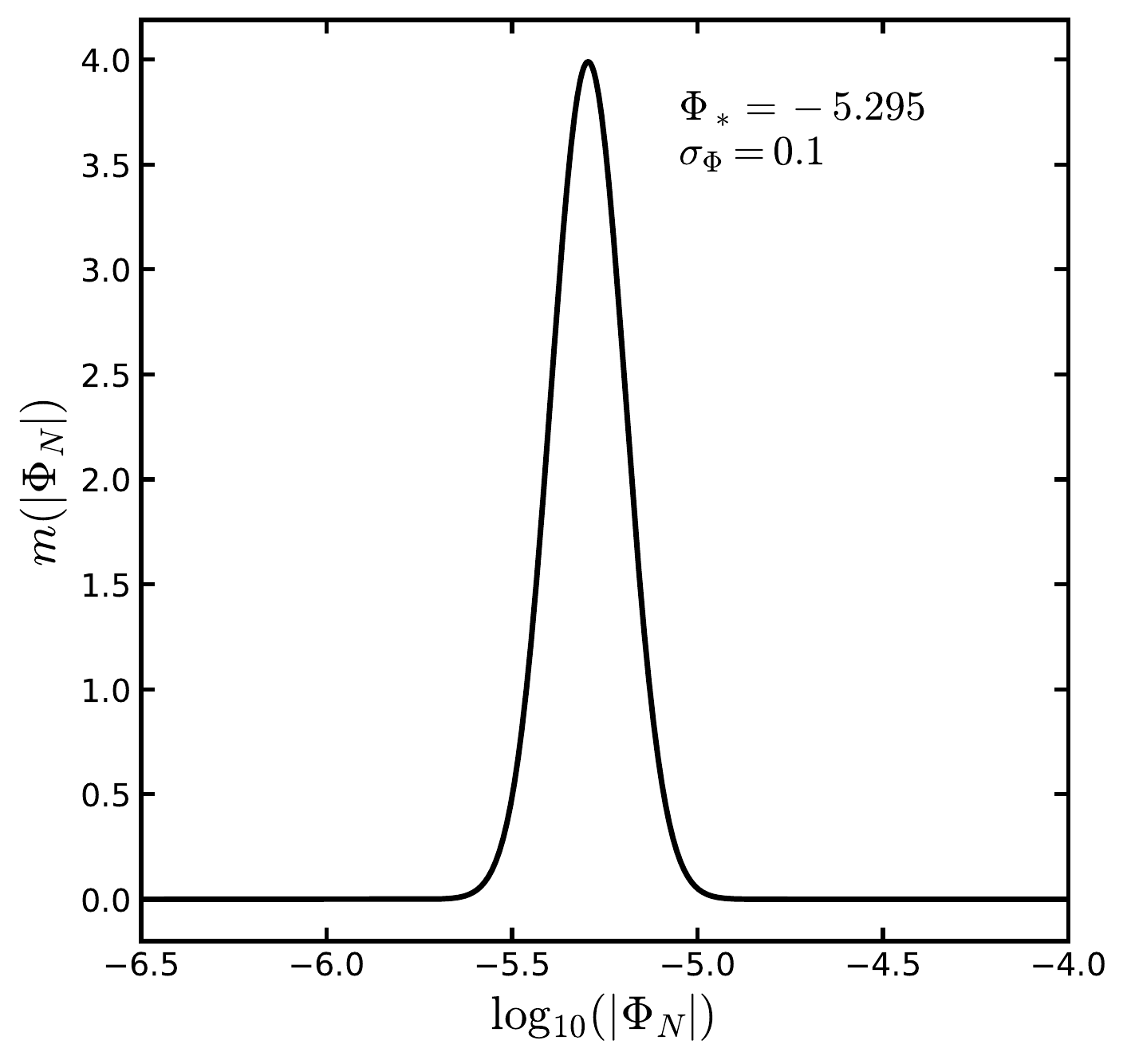}
\includegraphics[width=0.33\textwidth]{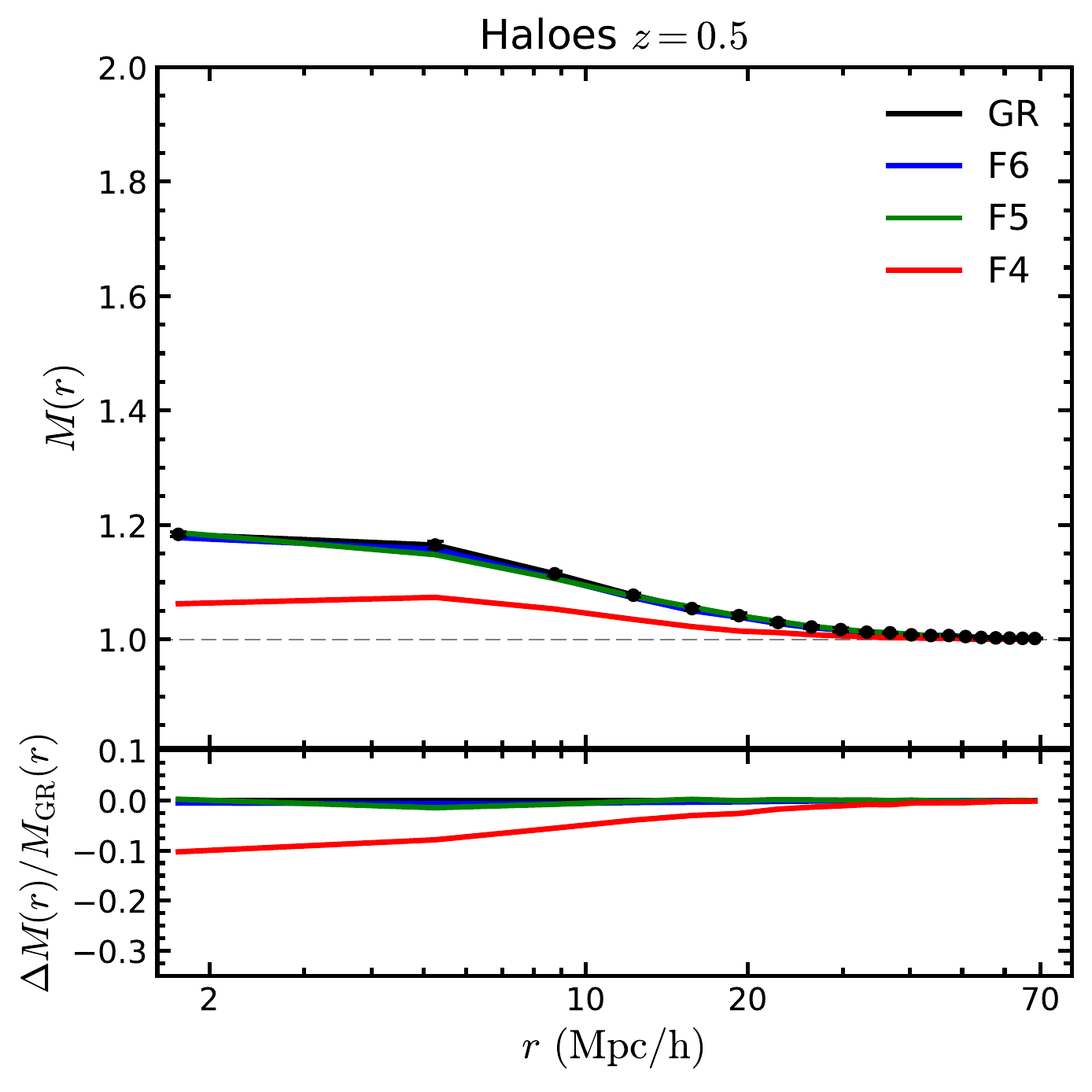}
\includegraphics[width=0.33\textwidth]{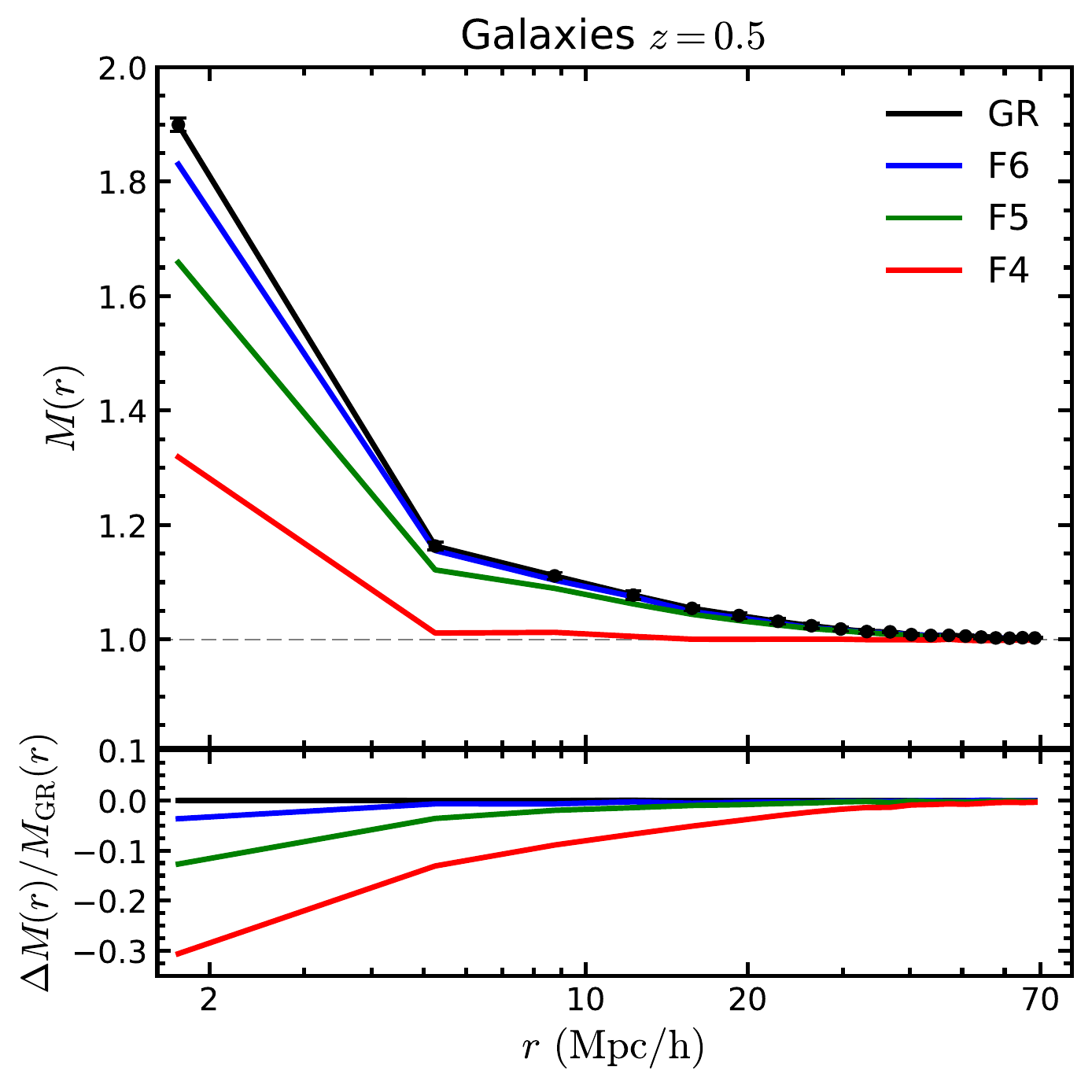}
	\caption{Marked correlation functions of haloes and CMASS galaxies at $z=0.5$; mark in function of the Newtonian gravitational potential. Left-hand side panel shows the functional form of the Gaussian-$\Phi_{\rm N}$ mark \eqref{eq:mGphi}; the values of the parameters $\Phi_*$ and $\sigma_\Phi$ are shown in the legend. Middle and right-hand side plots show the marked correlation function using the mark given by Eqn.~\eqref{eq:mGphi} for haloes and galaxies, respectively. All lower subpanels for middle and right-hand side plots show the relative difference between $f(R)$ models and GR. The plotted values correspond to the average over the 5 realisations. Errors correspond to $1\sigma$ standard deviation over the 5 GR realisations.}
	\label{fig:mCF_Phi}
\end{figure*}

\subsubsection{Gravitational potential}\label{sub:pot}

Our second definition of environment is based on the Newtonian gravitational potential produced by dark matter haloes. The dark matter haloes in our simulations are reasonably well described by a NFW density profile \citep{Navarro:1995iw,Navarro:1996gj}:
\begin{equation}\label{eq:rho_nfw}
\rho_\textrm{NFW} = \frac{\rho_{\rm s}}{(r/r_{\rm s})(1+r/r_{\rm s})^2}\, , 
\end{equation}
where $r_{\rm s}$ is the scale radius where the profile has a slope of $-2$ and $\rho_{\rm s}$ is the characteristic density.
The Newtonian gravitational potential is obtained by solving the Poisson equation, $\nabla^2 \Phi_{\rm N} = 4\pi G \rho_\textrm{NFW}$, for the NFW density profile Eqn.~\eqref{eq:rho_nfw} 
\citep*{Cole:1996nfw,Navarro:1996gj,Lokas:2000mu}:
\begin{equation}\label{eq:PhiN}
\Phi_{\rm N} = - \frac{G M_{200c}}{r_{200c}} \frac{\ln(1 + c)}{\ln(1 + c) - c/(1 +c)}\,,
\end{equation}
where $G$ is Newton's gravitational constant, $M_{200c}$ was defined in Eqn.~\eqref{eq:M200c} and $c$ is the concentration parameter defined as $c\equiv r_{200c}/r_{\rm s}$. Previous studies have used the Newtonian gravitational potential in modified gravity to characterise local variations in the strength of gravity (see e.g., \citealt{Cabre:2012tq,Stark:2016mrr,Shi:2017pyd}).

For this environment definition we define a Gaussian mark which allows us to up-weight galaxies in some regions of interest,
\begin{equation}\label{eq:mGphi}
m = \frac{1}{\sqrt{2\pi}\sigma_{\Phi}} \exp\left[- \frac{(\log_{10}(|\Phi_{\rm N}|) - \Phi_*)^2}{2\sigma_{\Phi}^2} \right]\,,
\end{equation}
where $\Phi_*$ and $\sigma_\Phi$ are free parameters of the mark which control the amplitude and width of the regions highlighted. As we can see from the distribution of galaxies as a function of host halo mass (right panel of Fig.~\ref{fig:HOD}), most galaxies live in haloes with masses between $10^{13} < M_{200c}/[h^{-1} M_\odot] < 10^{14}$ (which correspond to the mass range of groups of galaxies). Hence, we use the Gaussian-$\Phi_{\rm N}$ mark to up-weight galaxies contained in these haloes. In principle we should be able to find a bigger difference in the clustering between the GR and $f(R)$ models using this mark, as suggested by the cumulative halo mass function (lower panel of Fig.~\ref{fig:HMF}).

The value of the centre of the Gaussian is $\Phi_* = -5.295$. This value was found by computing the Newtonian gravitational potential for each galaxy in Box 1 for GR, then we pick the maximum value found for haloes with $M_{200c} = 10^{14} h^{-1} M_\odot$ $(\Phi_{\rm max})$ and the minimum value for haloes with $M_{200c} = 10^{13} h^{-1} M_\odot$ $(\Phi_{\rm min})$, finally we take $\Phi_* = (\Phi_{\rm max} + \Phi_{\rm min})/2$. We tried different values of the width, finding that $\sigma_\Phi = 0.1$ best ensures that we only up-weight galaxies in the haloes of interest. The functional form of the Gaussian-$\Phi_{\rm N}$ mark, Eqn.~\eqref{eq:mGphi}, is shown in the left-hand panel of Fig.~\ref{fig:mCF_Phi}. 

The halo and galaxy marked correlation function is presented in the middle and right panels of Fig.~\ref{fig:mCF_Phi}, respectively. The results can be summarised as follows:
\begin{itemize}
\item In the case of the halo/galaxy marked correlation function (middle and right panels of Fig.~\ref{fig:mCF_Phi}, respectively), the two-point correlation function (used as the denominator of Eqn.~\eqref{eq:Mr}) is lower than the weighted correlation function, leading to $M(r) \geq 1$ for all gravity models, due to the stronger clustering of the up-weighted haloes in the mass range $M_{200c}/[h^{-1} M_\odot] = [10^{13},10^{14}]$.
\item F6 predicts almost an identical halo/galaxy marked clustering to that in GR, which is consistent to our understanding that the screening mechanism in this model works efficiently in haloes of the mass range up-weighted.
\item For F4 haloes, the mCF is higher than the 2PCF for the reason given in the first bullet point above. However, in this model a larger fraction of haloes in the mass range $M_{200c}/[h^{-1} M_\odot] \in [10^{13},10^{14}]$ are formed from low initial density peaks (due to stronger gravity) which are not very strongly clustered, such that the up-weighting of them -- while making $M(r)\geq1$ -- does not lead to a $M(r)$ as large as in GR. This leads to $\Delta M(r)/M_{\rm GR}(r)<0$ for F4. For F5 haloes, the fifth force is strong enough to enhance their clustering, but not too strong to produce excessive merging, and so the up-weighting using the Gaussian mark increases the mCF as significantly as in GR.
\item For galaxies, a key difference from haloes is that a halo can host several galaxies while some haloes do not host galaxies at all. In F4 and F5, more relatively low initial density peaks have been promoted to the halo mass range $M_{200c}/[h^{-1} M_\odot] \in [10^{13},10^{14}]$ due to the enhanced gravity, and at the same time some high density peaks have grown out of this mass range. This means that if we up-weight galaxies whose host haloes are in this mass range, we end up with more central and fewer satellite galaxies, and more of them are hosted by haloes from lower initial density peaks. By the same reasoning as above, while we still have $M(r)>1$ for these models, it is smaller than in GR and F6. In particular, we have noticed that $\Delta M(r)/M_{\rm GR}(r)$ reaches $5\sim10$\% for F5 and $20\sim30$\% for F4 in $r=2\sim5h^{-1}$Mpc. These results are very stable, and change very little across the different simulation realisations.
\end{itemize}

Also we note that the differences between the $f(R)$ and GR models are boosted when we use additional information to the density field. This can be seen by comparing the right panels of Fig.~\ref{fig:mCF_rho} with the right panel of Fig.~\ref{fig:mCF_Phi}. The differences get larger in such cases because the galaxy density field and galaxy distribution have been tuned to match between the different models (see Sec.~\ref{sub:hod_c}). In all cases we observe that signals above $20~h^{-1}$ Mpc become identical between models. This is because the marked correlation function is the ratio of two correlation functions (see right hand expression of Eqn.~\eqref{eq:Mr}) and we have $\xi(r) \sim W(r)$ for $r>20~h^{-1}$ Mpc.

From the observational point of view, we can measure the Newtonian gravitational potential from the X-ray temperature of galaxy clusters (see e.g., \citealt{Allen:2004cd,Allen:2007ue,Li:2015rva}), the gas mass fractions of clusters and the escape velocity profile, $v_\mathrm{esc}(r)$ \citep{Stark:2016mrr}. Hence, if we reconstruct the gravitational potential from the observations mentioned above and use a mark that is a  function of the potential, similar to Eqn.~\eqref{eq:mGphi}, then we can test this approach and potentially find a measurable signature of modified gravity. One caveat is that the gravitational potential constructed in this way is the dynamical potential, while in this study we have used lensing potential of haloes \citep[see, e.g.,][]{He:2015mva}.

\section{Discussion and conclusions}\label{sec:conclusions}

We study the clustering of haloes and galaxies in four different cosmologies: a $\Lambda$CDM 
model which is based on general relativity and three \citeauthor{Hu:2007nk} $f(R)$ chameleon models with fixed $n=1$ and $|f_{R0}| = 10^{-6}, 10^{-5},10^{-4}$ (denoted F6, F5 and F4). We analyse the output of dark matter-only N-body simulations related to these models at $z=0.5$. 

First, we study the cumulative halo mass function, finding that the F4 model predicts more haloes than GR at all masses probed by our simulations, with the maximum difference reaching an excess of more than 50 percent for haloes with mass $M_{200c} > 10^{14.3} h^{-1} {\rm M_\odot}$. These differences occur due to the enhancement of gravity in $f(R)$ gravity, which results in the production of more massive haloes in F4 than GR through faster accretion and more frequent merging of small haloes. The differences found in F5 reach $25\%$ for haloes with masses $10^{13} h^{-1} {\rm M_\odot} < M_{200c} < 10^{14} h^{-1} {\rm M_\odot}$ where the screening mechanism at this mass scale is inefficient for this model. F6  shows the smallest difference from GR because in this model the chameleon screening is strong in haloes with mass $M_{200c} > 10^{13} h^{-1} {\rm M_\odot}$, thereby suppressing the effects  of the fifth force. 

We populate dark matter haloes with galaxies using a halo occupation distribution, using a five-parameter model which treats separately central and satellite galaxies, with the values of the parameters as used in  \citet{Manera:2012sc} to reproduce the clustering of CMASS galaxies with a density number, $n_{\rm g} = 3.2 \times 10^{-4} h^3 {\rm Mpc}^{-3}$ \citep{Anderson:2012sa} for our GR simulations. We tuned the parameters to match the galaxy number density and two-point correlation function of GR to within $1-3\%$ for the $f(R)$ models. 
The galaxy two-point correlation functions for the $f(R)$ and GR models are presented in the right plot of Fig.~\ref{fig:2pcf}.

Then we study the two-point clustering of dark matter haloes. We choose samples of haloes with fixed halo number density, $n_{\rm h} = n_{\rm g}$, resulting to different mass cutoffs in our halo catalogues for all gravity models: $7.643 \times 10^{12} h^{-1}M_\odot$ (GR), $7.798 \times 10^{12} h^{-1}M_\odot$ (F6), $9.124 \times 10^{12} h^{-1}M_\odot$ (F5) and  $8.734 \times 10^{12} h^{-1}M_\odot$ (F4). We find significant differences in the clustering of dark matter haloes for $f(R)$ models with respect to the GR predictions. The maximum difference between F4 and GR is $\sim20\%$, while for F5 and F6 it is less than $5\%$. Also we note that haloes in F5 are more clustered than those haloes in the $\Lambda$CDM model, whereas for F6 and F4 haloes are less clustered than their GR counterparts. These results are the effects of the enhancement of gravity which means a stronger growth of density peaks and therefore more massive structures at late times which gives a stronger clustering of the structures that are formed from these density peaks.  

To investigate whether or not these differences could be boosted by using an alternative approach to measure galaxy clustering, we used the marked correlation function \citep{Sheth:2005aj,White:2016yhs}. For this purpose we use two definitions for the environment of galaxies/haloes: a) the number density field and b) the Newtonian gravitational potential of the host halo. For the former we analyse three marks: i) an inverse power-law which enhances low-density regions (see Eqn.~\eqref{eq:W16}), ii) a log-transform mark which up-weighs intermediate- and high-density regions (see Eqn.~\eqref{eq:log}) and iii) a Gaussian mark given by Eqn.~\eqref{eq:mGphi} which allows us to up-weight only intermediate-density regions, $\rho_R = 1.25 - 1.88$. For the latter we use a Gaussian mark which allows us to up-weight haloes (and galaxies within those haloes) with mass $10^{13} < M_{200c}/[h^{-1} M_\odot] < 10^{14}$. 

We found that the halo and galaxy marked correlation functions for F6 is indistinguishable from GR using all marks, except for the galaxy Gaussian ($\rho_R$ and $\Phi_{\rm N}$) marked correlation functions which predict differences at most $\sim5\%$ from GR. 
For the F5 (F4) model, we notice that galaxies mimic the marked clustering at least for the White and log density-marks finding differences of $5\%$ $(2.5\%)$ and $2.5\%$ $(2.5\%)$, respectively. 
On the other hand, we observe that with the Gaussian marks (density field and gravitational potential) the difference in the galaxy marked correlation function is boosted, especially for F4, producing a difference of $20\%$ (using density field) and $30\%$ (using gravitational potential) with respect GR.

The galaxy marked correlation functions show smaller differences between the $f(R)$ and GR models for the density marks, Eqns.~\eqref{eq:W16}--\eqref{eq:gauss}, than in the case when using the gravitational potential mark, Eqn.~\eqref{eq:mGphi}, this is because the galaxy density field has been tuned to match between the different models. One caveat for our results is that there will be systematics when estimating the mark for observational samples. 

Another important feature we observe from marked correlation functions is that the signal above $20 \, h^{-1}$ Mpc does not distinguish between models (see corresponding plots of Figs.~\ref{fig:mCF_rho} and \ref{fig:mCF_Phi}). Instead, the measurable differences are on small scales. To improve our predictions on sub-Mpc scales we need to perform higher resolution  simulations, but we leave this for future work.

\cite{Valogiannis:2017yxm} recently found that using the dark matter distribution and the White-mark, Eqn.~\eqref{eq:W16} with $\rho_* = 4$ and $p=10$, the difference between F4 and GR marked correlation functions can reach a maximum of $37\%$ at $r=1.81 h^{-1}$ Mpc. These results can not readily be compared with ours, since we consider dark matter haloes and galaxies rather than the dark matter itself. Furthermore, we employ a different definition of density (counts-in-cells versus the cloud-in-cell smoothing used by Valogiannis \& Bean). Although their simulations are similar resolution to the ones we use, the volume of our boxes is $\sim 60$ times larger, which allows more robust clustering measurements. 

If we consider the statistical errors presented in this paper, then the differences between the $f(R)$ signals with respect GR are significant. It should be feasible to test the differentiating power of marked clustering statistics using real data from current galaxy surveys. In future work we need to improve the resolution of the simulations and make more realistic mock galaxy catalogues to allow a  fairer comparison with upcoming  observations.

Here we have demonstrated the potential of the marked correlation function to differentiate between gravity models. The next step is to extend these calculations, which were presented for massive galaxies, to the emission line galaxies that will be selected by the DESI and Euclid redshift surveys. 

\section*{Acknowledgements}
The authors wish to thank Lee Stothert for providing the code to compute the $2$-point correlation functions and the weighted number counts.
Also we would like to thank the referee for their positive and helpful comments about our paper.
CH-A is supported by the Mexican National Council of Science and Technology (CONACyT) through grant No. 286513/438352. BL is supported by the European Research Council (ERC-StG-716532-PUNCA). This project has received funding from the European Union's Horizon 2020 Research and Innovation Programme under the Marie Sk{\l}odowska-Curie grant agreement No 734374, LACEGAL. We acknowledge support from STFC Consolidated Grants ST/P000541/1, ST/L00075X/1. 
This work used the DiRAC Data Centric system at Durham University, operated by the Institute for Computational Cosmology on behalf of the STFC DiRAC HPC Facility (\url{www.dirac.ac.uk}). This equipment was funded by BIS National E-infrastructure capital grant ST/K00042X/1, STFC capital grants ST/H008519/1 and ST/K00087X/1, STFC DiRAC Operations grant ST/K003267/1 and Durham University. DiRAC is part of the National E-Infrastructure.




\bibliographystyle{mnras}
\bibliography{references} 






\bsp	
\label{lastpage}
\end{document}